\documentclass[twocolumn]{revtex4}
\usepackage{amsmath}
\usepackage[colorlinks, linkcolor=red,anchorcolor=green,citecolor=blue]{hyperref}
\begin{document}
\title{From Chiral Kinetic Theory To Relativistic Viscous Spin Hydrodynamics}

\author{Shuzhe Shi}
\author{Charles Gale} 
\author{Sangyong Jeon}

\address{Department of Physics, McGill University, 3600 University Street, Montreal, Quebec H3A 2T8, Canada.}
\begin{abstract}
In this work, we start with chiral kinetic theory and construct the spin hydrodynamic framework for a chiral spinor system. Using the 14-moment expansion formalism, we obtain the equations of motion of second-order dissipative relativistic fluid dynamics with non-trivial spin polarization density. In a chiral spinor system, the spin alignment effect could be treated in the same framework as the Chiral Vortical Effect (CVE). However, the quantum corrections due to fluid vorticity induce not only CVE terms in the vector/axial charge currents but also corrections to the stress tensor. In this framework, viscous corrections to the hadron spin polarization are self-consistently obtained, which will be important for precise prediction of the polarization rate for the observed hadrons, e.g. $\Lambda$-hyperon.
\end{abstract}
\maketitle

\section{Introduction}
Relativistic heavy-ion collisions provide a special environment to study the strong interaction. 
In such experiments, a new phase of matter --- the Quark-Gluon Plasma (QGP) --- is created~\cite{Adams:2005dq,Gyulassy:2004zy}. 
Recently the STAR Collaboration at the Relativistic Heavy Ion Collider (RHIC) reported measurement of a non-vanishing polarization of $\Lambda$-hyperons~\cite{STAR:2017ckg,Adam:2018ivw}.
This result could imply an extremely vortical fluid flow structure in the QGP produced in semi-central nucleus-nucleus collisions, and has attracted significant interest and generated wide enthusiasm. 
In addition, detailed measurement of the spin polarization, in particular the longitudinal polarization at different azimuthal angles~\cite{Adam:2019srw}, disagrees with current theoretical expectation~\cite{Becattini:2015ska,Becattini:2017gcx,Xia:2018tes}.

In theoretical attempts (e.g.~\cite{Becattini:2016gvu,Xie:2015xpa,Xie:2019jun,Shi:2017wpk,Guo:2019joy,Li:2017slc,Karpenko:2016jyx}) to compute the hadron polarization rate, one typically assumes that hadrons are created according to the thermal equilibrium distribution for particles in a locally rotating fluid, whereas the viscous corrections induced by off-equilibrium effects are neglected. Also, studies generally assume that the spin degrees of freedom of either hadrons or partons have negligible influences on the dynamical motion of the medium. A more sophisticated and self-consistent framework is required to understand the discrepancy alluded to above, and to describe the vortical structure of QGP. Consequently, we propose to develop a relativistic dissipative hydrodynamic theory with spin degrees of freedom, i.e. ``spin hydrodynamics'', from a microscopic theory with the vortical and non-equilibrium effects consistently taken into account~\cite{Shi:2020qrx}. As a first step, we concentrate on the chiral limit in this work, owing to its simple structure of the underlying microscopic theory.

Although hydrodynamics is a macroscopic theory based on conservation laws and the second law of thermodynamics, the evolution of dissipative quantities depends on the details of how the system approaches the thermal distribution and needs the guidelines of kinetic theory to correctly reflect microscopic processes.
In a massless fermion system, the microscopic transport processes are described by the Chiral Kinetic Theory (CKT)~\cite{Son:2012wh,Son:2012zy,Stephanov:2012ki}.
A convenient way to derive the CKT is the Wigner function formalism~\cite{DeGroot:1980dk,Vasak:1987um,Gao:2012ix,Hidaka:2016yjf,Gao:2017gfq,Liu:2018xip,Huang:2018wdl}.
For the pedagogical reason, we review recent developments in the Wigner function formalism of Chiral Kinetic Theory in Sec.~\ref{sec.ckt}. 
With such a tool, we derive ideal spin hydrodynamic equations for thermal equilibrium systems in Sec.~\ref{sec.equilibrium}, and obtain viscous spin hydrodynamics in Sec.~\ref{sec.non_equilibrium}. 
In addition, we analyze the causality and stability of spin hydrodynamic equations against linear perturbations in App.~\ref{sec.sound_mode}, and explore the pseudo-gauge transformation to symmetrize the stress-tensor in App.~\ref{sec.pseudogauge}.
In the rest of the appendices, we include calculation details.

In this paper, we take the mostly-negative convention of metric $g^{\mu\nu}=\mathrm{diag}(+,-,-,-)$, and adopt the following notation:
\begin{eqnarray}
&&	\Delta^{\mu\nu} \equiv g^{\mu\nu} - u^\mu u^\nu \,,\\
&&	\Delta^{\mu\nu}_{\alpha\beta} \equiv \frac{1}{2}\Delta^{\mu}_{\alpha}\Delta^{\nu}_{\beta} + \frac{1}{2}\Delta^{\mu}_{\beta}\Delta^{\nu}_{\alpha} - \frac{1}{3}\Delta^{\mu\nu}\Delta_{\alpha\beta} \,,\\
&&	\varpi_{\mu\nu} \equiv \frac{1}{2}\Big(\partial_\nu \frac{u_\mu}{T} - \partial_\mu \frac{u_\nu}{T} \Big) \,,\\
&&	\omega^\mu \equiv -\frac{T}{2}\epsilon^{\mu\nu\rho\sigma}u_\nu \varpi_{\rho\sigma} = \frac{1}{2}\epsilon^{\mu\nu\rho\sigma}u_\nu \partial_\rho u_\sigma \,,\\
&&	\hat{\mathrm{d}} X \equiv u^\mu \partial_\mu X \,,\\
&&	\theta \equiv \partial_\mu u^\mu \,,\\
&&	\sigma^{\mu\nu} \equiv \Delta^{\mu\nu}_{\alpha\beta} \partial^\alpha u^\beta \,.
\end{eqnarray}
In addition, we define the projected vector/tensor as:
\begin{eqnarray}
&&	V^{\langle \alpha \rangle} \equiv \Delta^{\alpha}_{\mu} V^\mu \,,\\ 
&&	V^{\langle \alpha \beta \rangle} \equiv \Delta^{\alpha\beta}_{\mu\nu} V^{\mu\nu} \,,\\ 
&&	V^{\langle \alpha} U^{\beta \rangle} \equiv \Delta^{\alpha\beta}_{\mu\nu} V^{\mu}U^{\nu} \,.
\end{eqnarray}

\section{Chiral Kinetic Theory from Wigner Function Formalism}\label{sec.ckt}
Spin is an intrinsic quantum degree of freedom of elementary particles. 
To describe the non-equilibrium collective behavior of Dirac spinors taking into account the spin degrees of freedom, a natural framework is the Wigner formalism:
\begin{equation}
W_{ab}(x,p) \equiv \left< \int \mathrm{d}^4y\, e^{\frac{i}{\hbar}p\cdot y} \widehat{\bar\psi}_{b}(x+\frac{y}{2})\widehat{\psi}_{a}(x-\frac{y}{2}) \right> \,.
\end{equation}
As a $4\times4$ matrix depending on coordinate $x$ and momentum $p$, it describes the phase space distribution for different spin states, and can be decomposed in the Clifford basis $\{ I, \gamma^\mu, \gamma^5\equiv i \gamma^0 \gamma^1 \gamma^2 \gamma^3, \gamma^5 \gamma^\mu, \Sigma^{\mu\nu} \equiv \frac{i}{2}[\gamma^\mu,\gamma^\nu]\}$,
\begin{eqnarray}
W \equiv 
	\frac{1}{4}\Big( \mathcal{F} + i\mathcal{P} \gamma^5 + \mathcal{V}_\mu \gamma^\mu 
	+  \mathcal{A}_\mu \gamma^5 \gamma^\mu + \frac{1}{2} \mathcal{L}_{\mu\nu}\Sigma^{\mu\nu} \Big),\;\;
\end{eqnarray}
where the scalar $\mathcal{F}$, pseudo-scalar $\mathcal{P}$, vector $\mathcal{V}_\mu$, axial-vector $\mathcal{A}_\mu$, and tensor $\mathcal{L}_{\mu\nu}$ are known as the Clifford components.
With these components, one can express the thermodynamic quantities --- the current, axial current, energy-momentum tensor, and the spin tensor current --- respectively, as
\begin{eqnarray}
J^{\mu} &\equiv& \langle \bar\psi \gamma^\mu \psi \rangle = \int \frac{\mathrm{d}^4p}{(2\pi)^4} \mathcal{V}^\mu \,,\label{eq.current}\\
J^{\mu}_{A} &\equiv& \langle \bar\psi \gamma^\mu\gamma^5 \psi \rangle = \int \frac{\mathrm{d}^4p}{(2\pi)^4} \mathcal{A}^\mu \,,\label{eq.axial}\\
T^{\mu\nu} &\equiv& \langle \bar\psi (i\gamma^\mu D^\nu) \psi \rangle = \int \frac{\mathrm{d}^4p}{(2\pi)^4} p^\nu \mathcal{V}^\mu \,,\label{eq.stresstensor}\\
S^{\lambda\mu\nu} &\equiv& \frac{1}{8} \langle \bar\psi \{\gamma^\lambda , \Sigma^{\mu\nu} \} \psi \rangle
	= \frac{1}{2}\epsilon^{\sigma\lambda\mu\nu} \int \frac{\mathrm{d}^4p}{(2\pi)^4} \mathcal{A}_\sigma \,. \quad\label{eq.spin}
\end{eqnarray}

In the absence of an external field, the equation of motion for the Wigner function can be obtained from the Dirac equation
\begin{equation}
\gamma_\mu (p^\mu + \frac{1}{2}i\hbar \partial^\mu) W(x,p) = m\, W(x,p)\,,
\label{eq.Wigner_equation}
\end{equation}
which contains a set of coupled equations for the Clifford components.
In {\it the massless limit} $(m=0)$, the equations are partially decoupled and the vector and axial-vector components, $\mathcal{V}_{\mu}$ and $\mathcal{A}_{\mu}$, couple only with each other, but not the scalar, pseudo-scalar, and tensor components
\begin{eqnarray}
&p^{\mu}\mathcal{V}_{\mu}=0,\qquad 
	p^{\mu}\mathcal{A}_{\mu}=0,\\
&\partial^{\mu}\mathcal{V}_{\mu}=0,\qquad
	\partial^{\mu}\mathcal{A}_{\mu}=0, \\
&\frac{\hbar}{2}\epsilon_{\mu\nu\rho\sigma}\partial^{\rho}\mathcal{V}^{\sigma}
	= p_{\nu}\mathcal{A}_{\mu}-p_{\mu}\mathcal{A}_{\nu}, \\
&\frac{\hbar}{2}\epsilon_{\mu\nu\rho\sigma}\partial^{\rho}\mathcal{A}^{\sigma}
	= p_{\nu}\mathcal{V}_{\mu}-p_{\mu}\mathcal{V}_{\nu}. \label{eq.angularmomentum_conservation}
\end{eqnarray}
These equations can be further simplified by re-combining vector and axial-vector into left-handed (LH) and right-handed (RH) components, $\mathcal{J}_\pm^\mu \equiv \frac{1}{2}(\mathcal{V}^\mu \pm \mathcal{A}^\mu)$. 
They evolve independently
\begin{eqnarray}
&&p^{\mu}\mathcal{J}_{\pm,\mu}=0,\\
&&\partial^{\mu}\mathcal{J}_{\pm,\mu}=0, \\
&&\frac{\hbar}{2}\epsilon_{\mu\nu\rho\sigma}\partial^{\rho}\mathcal{J}^{\sigma}_\pm
	= \pm( p_{\nu}\mathcal{J}_{\pm,\mu}-p_{\mu}\mathcal{J}_{\pm,\nu} ).
\end{eqnarray}
In Refs.~\cite{Hidaka:2016yjf,Huang:2018wdl}, the authors employ a semi-classical expansion (i.e. $\hbar$ expansion) in the massless limit, and derive the CKT up to the leading order in $\hbar$.
In first order CKT, the RH/LH components can be expressed as:
\begin{equation} \label{eq.chiral-current}
\mathcal{J}_\pm^\mu
	= \Big( p^\mu \pm \hbar \frac{\epsilon^{\mu\nu\rho\sigma}p_\rho n_\sigma}{2 p \cdot n} \partial_\nu \Big) f_\pm \,,
\end{equation}
where $f_\pm$ are the RH/LH particle distribution functions, defined as the $p^\mu$-proportional section of corresponding chirality current $\mathcal{J}_\pm^\mu$. 
Their equations of motion are driven by the Chiral Kinetic Equation (CKE):
\begin{equation} \label{eq.cke}
\Big[p^\mu \partial_\mu \pm
	\hbar \Big(\partial_\mu \frac{\epsilon^{\mu\nu\rho\sigma}p_\rho n_\sigma}{2 p \cdot n}\Big)\partial_\nu
	\Big]f_\pm = 0 \,.
\end{equation}
In particular, $n^\mu$ is a time-like arbitrary auxiliary vector field, and could depend on space-time $x^\mu$ in a non-trivial way.
It is introduced to separate the $p^\mu$-parallel and $p^\mu$-perpendicular components.
Noting that the momentum $p^\mu$ is a null vector hence self-perpendicular, the separation is not unique and depends on the choice of $n^\mu$.
Such non-uniqueness leads to the frame-dependence of the distribution function --- also known the {\it side-jump effect}~\cite{Chen:2015gta,Hidaka:2016yjf,Huang:2018wdl}.
When choosing different auxiliary field, e.g. $u^\mu$ and $v^\mu$, the corresponding distribution functions, $f_{[u],\pm}$ and $f_{[v],\pm}$, differ at $\hbar$-order:
\begin{equation}
f_{[u],\pm} - f_{[v],\pm} = \mp \hbar \frac{\epsilon^{\mu\nu\rho\sigma}p_\mu u_\nu v_\rho \partial_\sigma f_{(0)\pm}}{2(u\cdot p)(v \cdot p)}\,,
\end{equation}
and consequently
\begin{equation}
p^\mu f_{[u],\pm} - p^\mu  f_{[v],\pm}
	= \mp  \hbar \Big( \frac{\epsilon^{\mu\nu\rho\sigma}p_\rho u_\sigma}{2 p \cdot u}
		- \frac{\epsilon^{\mu\nu\rho\sigma}p_\rho v_\sigma}{2 p \cdot v} \Big)  \partial_\nu  f_{(0),\pm} \,,
\end{equation}
so that the definition of $\mathcal{J}_\pm^\mu$ remains invariant. We refer the readers to~\cite{Huang:2018wdl} for detailed derivations. 
In the above equations, $f_{(0)\pm}$ is the classical $\hbar^0$-order of chirality density function and is frame-independent.
As will be discussed in Sec.~\ref{subsec.ideal_hydro}, it will be more natural to choose $n^\mu$ the local fluid velocity. For the sake of generality, we keep $n^\mu$ to be arbitrary at this point.

Last but not least, the conservation equation of total angular momentum
\begin{align}\begin{split}
0 =&\; \partial_\mu \mathcal{M}^{\mu\nu\lambda} \\
\equiv&\; \partial_\mu (\mathcal{L}^{\mu\nu\lambda} + \hbar{S}^{\mu\nu\lambda}) \\
\equiv&\;  \partial_\mu( T^{\mu\lambda}x^{\nu} - T^{\mu\nu}x^{\lambda}) + \hbar\partial_\mu {S}^{\mu\nu\lambda} \\
=&\; (T^{\nu\lambda} - T^{\lambda\nu}) + \hbar \partial_\mu {S}^{\mu\nu\lambda} \,,
\end{split}\end{align}
is satisfied automatically, which can be shown by taking the momentum integral of Eq.~\eqref{eq.angularmomentum_conservation}, one of the equations of motion for Wigner components.
In a system with Dirac spinors, the conservation of total angular momentum is not an extra constraint on the system evolution.
The spin density current follows once the axial charge
density, accounting for the imbalance between right-handed (RH) and left-handed (LH) particles,
is defined.

\section{Spin Hydrodynamics in Equilibrium}\label{sec.equilibrium}
\subsection{Equilibrium Distribution}\label{sec.equilibrium_distribution}
To connect kinetic theory with hydrodynamic theory, a natural starting point is the equilibrium limit of the distribution function. 
This is non-trivial when rotation effects are included: quantum corrections appear in the kinetic equation Eq.\eqref{eq.cke}, therefore the equilibrium distribution will also be modified.
Here we derive the equilibrium distribution with vorticity corrections, $f_{\pm,\mathrm{eq}}$, in a similar way as in Ref.~\cite{Liu:2018xip}.
We start from the principle that equilibrium distribution $f_\pm(x,p) \equiv f_{\pm}(g_\pm)$ should be a function of the linear combination of the quantities conserved in collisions --- namely, the particle number, the momentum, and the angular momentum, 
\begin{equation}
g_\pm = \alpha_\pm + \beta_\lambda p^\lambda + \hbar \gamma_{\pm,\mu\nu} \frac{\epsilon^{\mu\nu\alpha\beta}p_\alpha n_\beta}{2 p \cdot n},
\end{equation}
where the coefficient $\alpha$, $\beta$, and $\gamma$ are not arbitrary.
They are constrained by the CKE:
\begin{equation}\begin{split}
0=\,&\delta(p^2) \Big[p^\mu \partial_\mu 
	\pm\hbar \Big(\partial_\mu \frac{\epsilon^{\mu\nu\rho\sigma}p_\rho n_\sigma}{2 p \cdot n}\Big)\partial_\nu
	\Big]f_\pm(g_\pm) \\
=\,&\delta(p^2) \frac{\mathrm{d}f_\pm}{\mathrm{d}g_{\pm}}\Big[p^\mu \partial_\mu 
	\pm\hbar \Big(\partial_\mu \frac{\epsilon^{\mu\nu\rho\sigma}p_\rho n_\sigma}{2 p \cdot n}\Big)\partial_\nu
	\Big]g_\pm \,.
\end{split}\end{equation}
To solve the coefficients, we take the semi-classical expansion
\begin{equation}\begin{split}
g_\pm 
=\,&	g_{(0),\pm} + \hbar g_{(1),\pm} + \mathcal{O}(\hbar^2) \\
=\,&	\Big( \alpha_{(0),\pm} + p_\mu \beta_{(0)}^\mu \Big) 
	+ \hbar \Big( \alpha_{(1),\pm} + p_\mu \beta_{(1)}^\mu 
\\&	+ \gamma_{\pm,\mu\nu}  \frac{\epsilon^{\mu\nu\alpha\beta}p_\alpha n_\beta}{2 p \cdot n}\Big)  
	+  \mathcal{O}(\hbar^2) \,,
\end{split}\end{equation}
as well as
\begin{equation}\begin{split}
f_\pm(g_\pm) 
=\,&	f_{(0),\pm}(g_{(0),\pm}) 
	+ \hbar f'_{(0),\pm}(g_{(0),\pm}) \Big( \alpha_{(1),\pm} 
\\&	 + p_\mu \beta_{(1)}^\mu + \gamma_{\pm,\mu\nu}  \frac{\epsilon^{\mu\nu\alpha\beta}p_\alpha n_\beta}{2 p \cdot n}\Big) 
	+  \mathcal{O}(\hbar^2) \,.
\end{split}\end{equation}
From zeroth order CKE, one finds that 
\begin{equation}
	\partial_\mu \alpha_{(0),\pm} = 0,\quad
	\partial_\mu \beta_{(0),\nu} + \partial_\nu \beta_{(0),\mu} = \frac{\partial \cdot \beta_{(0)}}{4} g_{\mu\nu} \,.\label{eq.alpha_beta_0}
\end{equation}
Noting that $n^\mu$ is the auxiliary vector in constructing the solution of the Wigner function, one would need to ensure that physical quantities like $\mathcal{J}_\pm^\mu$ shall be independent of $n^\mu$, hence
\begin{equation}\begin{split}
& f_{[u],\pm} - f_{[v],\pm} 
\\=& 	\mp \hbar \frac{\epsilon^{\mu\nu\rho\sigma}p_\mu u_\nu v_\rho \partial_\sigma f_{(0)\pm}}{2(u\cdot p)(v \cdot p)} 
\\=&	\mp \hbar \frac{\epsilon^{\mu\nu\rho\sigma}p_\mu u_\nu v_\rho \partial_\sigma g_{(0)\pm}}{2(u\cdot p)(v \cdot p)}
	f'_{(0),\pm}(g_{(0),\pm}) + \mathcal{O}(\hbar^2)\,.
\label{eq.frame_dependence}
\end{split}\end{equation}
Comparing the above two equalities, one obtains that
\begin{equation}\begin{split}
&	\mp \frac{p^\lambda\epsilon^{\mu\nu\rho\sigma}p_\mu u_\nu v_\rho}{2(u\cdot p)(v \cdot p)} \partial_\sigma \beta_{(0),\lambda}
\\=\,&
	\gamma_{\pm,\mu\nu} \Big( \frac{\epsilon^{\mu\nu\alpha\beta}p_\alpha u_\beta}{2 p \cdot u} 
	-  \frac{\epsilon^{\mu\nu\alpha\beta}p_\alpha v_\beta}{2 p \cdot v} \Big)
\\=\,&
	2\gamma_{\pm,\lambda\sigma} \frac{p^\lambda \epsilon^{\mu\nu\rho\sigma}p_\mu u_\nu v_\rho }{2(u\cdot p)(v \cdot p)} \,.
\end{split}\end{equation}
Further noting the arbitrariness of $u$, $v$, and $p$, one gets 
\begin{equation}
\gamma_{\pm,\mu\nu}  = \pm \frac{1}{4} \Big(\partial_\mu \beta_{(0),\nu} - \partial_\nu \beta_{(0),\mu}\Big)\,.
\end{equation}
Then we consider the first order CKE and find
\begin{equation}
	\partial_\mu \alpha_{(1),\pm} = 0,\quad
	\partial_\mu \beta_{(1),\nu} + \partial_\nu \beta_{(1),\mu} = \frac{\partial \cdot \beta_{(1)}}{4} g_{\mu\nu} \,.\label{eq.alpha_beta_1}
\end{equation}
Consequently, one can absorb $\alpha_{(1),\pm}$ and $\beta_{(1),\mu}$, respectively, into $\alpha_{(0),\pm}$ and $\beta_{(0),\mu}$, and conclude that
\begin{equation}\begin{split}
&	\partial_\mu \alpha_\pm = 0,\qquad
	\partial_\mu \beta_\nu + \partial_\nu \beta_\mu = \frac{\partial \cdot \beta}{4} g_{\mu\nu} ,
\\&	\gamma_{\pm}^{\mu\nu} = \pm \frac{1}{4}(\partial^\mu \beta^\nu- \partial^\nu \beta^\mu)\,,
\end{split}\end{equation}
and 
\begin{equation}\begin{split}
f_\pm(g_\pm) 
=\,& f_{\pm}(\alpha_{\pm} + p_\mu \beta^\mu)
	 \pm \hbar \Big(\frac{\partial_\mu \beta_\nu- \partial_\nu \beta_\mu}{4}\times
\\&
	 \frac{\epsilon^{\mu\nu\alpha\beta}p_\alpha n_\beta}{2 p \cdot n}\Big) f'_{\pm}(\alpha_{\pm} + p_\mu \beta^\mu)
		+  \mathcal{O}(\hbar^2) \,.
\end{split}\end{equation}
It is worth noting that compared to the derivation in Ref.~\cite{Liu:2018xip}, we take into account the guiding principle that physical quantities are independent of $n^\mu$, i.e. Eq.~\eqref{eq.frame_dependence}.
By doing this, one would be able to rule out the ambiguous extra mode of $\gamma_{\pm,\mu\nu}$ pointed out in~\cite{Liu:2018xip}. 
Additionally, the conditions \eqref{eq.alpha_beta_0} and \eqref{eq.alpha_beta_1} apply only for a system in global equilibrium. They are not required in the derivation of the hydrodynamic equations.

Comparing the general form with momentum-integrated thermodynamics quantities, one can find that $\alpha_\pm=\mu_\pm/T$ corresponds to the RH/LH chemical potential, while $\beta_\mu\equiv u^\mu/T$ corresponds to the flow velocity and temperature. Particularly, the latter is independent of flavor or helicity.
Combined with the Fermi-Dirac distribution, we can express the equilibrium distribution functions in a compact form:
\begin{equation} \label{eq.thermal}
f_{\mathrm{eq},\pm}(p) 
	= \frac{1}{\exp[\frac{p\cdot u - \mu_\pm}{T} \mp \hbar \frac{\epsilon^{\mu\nu\rho\sigma}\varpi_{\mu\nu}p_\rho n_\sigma}{4\,n\cdot p}]+1} \,,
\end{equation}
where $\varpi_{\mu\nu} \equiv \frac{1}{2}\Big(\partial_\nu \frac{u_\mu}{T} - \partial_\mu \frac{u_\nu}{T} \Big)$ is the thermal vorticity.

\subsection{Ideal Spin Hydrodynamics}\label{subsec.ideal_hydro}
With the thermal distribution obtained, now we move on to construct the hydrodynamic quantities by taking the equilibrium limit.
For later convenience, we define the vorticity vector $\omega^\mu \equiv -\frac{T}{2}\epsilon^{\mu\nu\rho\sigma}u_\nu \varpi_{\rho\sigma} = \frac{1}{2}\epsilon^{\mu\nu\rho\sigma}u_\nu \partial_\rho u_\sigma$, the vector/axial chemical potential $\mu_V \equiv (\mu_++\mu_-)/2$, $\mu_A \equiv (\mu_+-\mu_-)/2$, 
and denote the integral $\int_p \equiv  \int \frac{2\delta(p^2)\mathrm{d}^4p}{(2\pi)^3}$. 
By substituting equilibrium distribution in the definition, the equilibrium hydrodynamic quantities are as follows:
\begin{eqnarray}
J_{\mathrm{eq},\pm}^{\mu} 
&\equiv&	\int_p  p^\mu f_{\mathrm{eq},\pm} 
	\pm \frac{\hbar}{2} \epsilon^{\mu\lambda\sigma\rho} 
	\int_p  \frac{p_\lambda n_\sigma}{n \cdot p} \partial_\rho f_{\mathrm{eq},\pm} 
\nonumber\\&=&
 	n_\pm u^\mu \pm \frac{\hbar}{2} \bigg(\frac{\partial n_\pm}{\partial \mu_\pm}\bigg)_{T,\mu_\mp} \omega^\mu
	\,,\\
J_{\mathrm{eq},V}^{\mu} 
&\equiv& J_{\mathrm{eq},+}^{\mu} + J_{\mathrm{eq},-}^{\mu} 
=
	n_V u^\mu + \frac{\hbar}{2} \bigg(\frac{\partial n_A}{\partial \mu_V}\bigg)_{T,\mu_A} \omega^\mu
	\,,\\
J_{\mathrm{eq},A}^{\mu} 
&\equiv& J_{\mathrm{eq},+}^{\mu} - J_{\mathrm{eq},-}^{\mu} 
=	n_A u^\mu + \frac{\hbar}{2} \bigg(\frac{\partial n_A}{\partial \mu_A}\bigg)_{T,\mu_V} \omega^\mu
	\,,\\
T_{\mathrm{eq}}^{\mu\nu} 
&\equiv&	
	\int_p p^\mu p^\nu (f_{\mathrm{eq},+} + f_{\mathrm{eq},-})
\nonumber\\&&
	+ \hbar \epsilon^{\mu\lambda\sigma\rho} 
	\int_p \frac{p^\nu p_\lambda n_\sigma}{2\;n \cdot p} \partial_\rho (f_{\mathrm{eq},+} - f_{\mathrm{eq},-}) 
\nonumber\\&=& 
	\varepsilon\, u^\mu u^\nu - P\,\Delta^{\mu\nu} + \frac{\hbar \,n_A}{4} (8 \omega^\mu u^\nu + T \epsilon^{\mu\nu\sigma\lambda}\varpi_{\sigma\lambda})
	\,,\nonumber\\~\\
S_{\mathrm{eq}}^{\lambda\mu\nu}
	&\equiv&  \frac{1}{2}\epsilon^{\lambda\mu\nu\sigma}J_{\mathrm{eq},A,\sigma} 
\end{eqnarray}
where
\begin{equation}
n_\pm \equiv \int_p  (u\cdot p) f_{\mathrm{eq},\pm}  \,,\qquad
\varepsilon = 3P \equiv \int_p  (u\cdot p)^2 (f_{\mathrm{eq},+} + f_{\mathrm{eq},-}) \,.
\end{equation}
We note that these are equivalent to the result in Ref.~\cite{Yang:2018lew}, if implementing the equilibrium distribution for both particle and anti-particle
\begin{equation}\begin{split}
n_\pm =\,& 
	\frac{\mu_\pm}{6} \Big(T^2 + \frac{\mu_\pm^2}{\pi^2}\Big) \,,\\
\varepsilon =\,& 
	\frac{7 \pi^2 T^4}{60} + \frac{T^2(\mu_V^2+\mu_A^2)}{2} + \frac{\mu_V^4 + 6\mu_V^2 \mu_A^2 + \mu_A^4}{4\pi^2}\,.
\end{split}\end{equation}

Some comments are in order:\\
(a) In the above equations, the quantum corrections to the vector and axial currents are collectively known as the Chiral Vortical Effect, (see e.g.~\cite{Son:2009tf}). 
In particular, even in the purely neutral case $\mu_V = \mu_A = 0$, the quantum correction to the axial current, $\hbar( T^2 \omega^\mu/6)$, is non-vanishing.
Noting that this leads to non-zero spin density $u_\lambda S_{\mathrm{eq}}^{\lambda\mu\nu} = \hbar T^3 \varpi^{\mu\nu}/12$, such a quantum correction term induces the spin-vorticity alignment.\\
(b) On top of an existing chiral-hydro that includes anomalous transport terms in the current and axial current, our derivation also indicates different terms in the stress tensor accounting for the feedback to energy and momentum flow. 
Quantum correction introduces an anti-symmetric term $\propto \hbar (4\omega^\mu u^\nu - 4\omega^\nu u^\mu + T \epsilon^{\mu\nu\sigma\lambda}\varpi_{\sigma\lambda})$,
together with a symmetric correction $\propto 4\hbar ( \omega^\mu u^\nu + \omega^\nu u^\mu)$. 
These terms are proportional to chirality imbalance, and vanish if $\mu_A = 0$, i.e. an equal amount of RH and LH particles at any spatial and temporal points.\\
(c) As a first-order derivative term $\omega^\mu$ appears in the hydrodynamic equations, it is non-trivial to show their causality and stability.
With details in App.~\ref{sec.sound_mode}, these equations are shown to be causal and stable against linear perturbations, which follow from the fact that $\partial_\mu \omega^\mu =(1/2)\epsilon^{\mu\nu\rho\sigma}(\partial_\mu u_\nu) (\partial_\rho u_\sigma)$ does not contain second-order derivatives of the velocity, such as $\partial_\alpha \partial_\beta u^\mu$.\\
(d) It might be worth noting that we take the canonical definition of energy-momentum tensor $T^{\mu\nu}$ and spin density $S^{\lambda\mu\nu}$. There have been discussions on the equivalence of evolution equations when taking other definitions, differing by a {\it pseudo-gauge transformation}~\cite{Becattini:2012pp,Florkowski:2018fap,Florkowski:2018ahw}.
In App.~\ref{sec.pseudogauge}, we derive the explicit form of the pseudo-gauge transformation to symmetrize $T^{\mu\nu}$. We emphasize that such a pseudo-gauge transformation does not cause an ambiguity, as the microscopic distribution, $f^\pm(p)$, is invariant under such a transformation. Physical observables, including the spin polarization vector, are constructed based on the distribution functions, hence they are not influenced by the pseudo-gauge transformation.\\
(e) Last but not the least, one can find that all these hydrodynamic quantities are independent of the choice of auxiliary field $n$, but the distribution functions, $f^\pm$, depends on the explicit form of $n^\mu$. We obtain the physical choice of such an auxiliary field as follows. We denote the spin correction term in distribution function \eqref{eq.thermal} as
\begin{equation}
\Sigma^{\mu\nu}_{[n]} \equiv \frac{\epsilon^{\mu\nu\alpha\beta}p_\alpha n_\beta}{2 n \cdot p}\,.
\end{equation} 
Noting that $n_\mu \Sigma^{\mu\nu}_{[n]} = 0$ transforms as a vector under Lorentz transformation, $\Sigma'^{\mu\nu}_{[n]}=\frac{\epsilon^{\mu\nu\rho0}p'_\rho}{2 E'_p}$ only contains the spatial part in the frame satisfying $n'^\mu = \{1,0,0,0\}$ at space-time point $(t',x',y',z')$. 
It represents the polarization tensor $\epsilon^{ijk}\hat{p}^k/2$ for a RH particle, whereas for a LH particle, the polarization tensor is $-\epsilon^{ijk}\hat{p}^k/2$, which is accounted for by the sign difference in the current term and equilibrium distribution function.
Consequently,  $\Sigma^{\mu\nu}_{[n]}$ serves as the spin tensor in the frame co-moving with $n^\mu$.
To correctly reflect the spin polarization in the distribution function, it is more natural to take $n^\mu = u^\mu$ to be the flow velocity. 
We adopt this choice for the rest of this paper.

\section{Hydrodynamics near Equilibrium}\label{sec.non_equilibrium}
In this section we extend the discussion to non-equilibrium systems, and derive second order spin hydrodynamics from the CKT.
To describe non-equilibrium hydrodynamics evolution, we start with the chiral kinetic equations with collision terms.
The quantum correction term in the CKE could be further simplified, see Eq.~\eqref{eq.quantum_correction_in_CKE} in App.~\ref{sec.mathematical_relations}.
Taking $n^\mu = u^\mu$,  the equations become
\begin{equation}
p^\mu \partial_\mu f_\pm
	\pm \hbar \Big(\frac{\epsilon^{\mu\nu\rho\sigma} p_\nu (\partial_\rho  u_\sigma)}{4\, u \cdot p}\Big)\partial_\mu f_\pm
	 = \mathcal{C}_\pm[f_+, f_-] \,,\label{eq.f_eom}
\end{equation}
where 
\begin{eqnarray}
\begin{split}
 \mathcal{C}_+(p) 
 =& \int_{\mathbf{k}, \mathbf{p'}, \mathbf{k'}} \Big[ 
 	W_{1}\Big(\tilde{f}_+(p') \tilde{f}_+(k') f_+(p) f_+(k) \\
&\qquad	- \tilde{f}_+(p) \tilde{f}_+(k) f_+(p') f_+(k')\Big) \\ 
&	\;\, +W_{2}\Big(\tilde{f}_+(p') \tilde{f}_-(k') f_+(p) f_-(k) \\
&\qquad 	- \tilde{f}_+(p) \tilde{f}_-(k) f_+(p') f_-(k')\Big) 
	\Big] \,,
\end{split}\\
\begin{split}
 \mathcal{C}_-(p) 
 =&	\int_{\mathbf{k}, \mathbf{p'}, \mathbf{k'}}  \Big[
  	W_{1}\Big(\tilde{f}_-(p') \tilde{f}_-(k') f_-(p) f_-(k) \\
&\qquad	- \tilde{f}_-(p) \tilde{f}_-(k) f_-(p') f_-(k')\Big) \\ 
&	\;\, +W_{2}\Big(\tilde{f}_-(p') \tilde{f}_+(k') f_-(p) f_+(k) \\
&\qquad - \tilde{f}_-(p) \tilde{f}_+(k) f_-(p') f_+(k')\Big) 
	\Big] \,,
\end{split}
\end{eqnarray}
are the collision kernels.
For later convenience, we recast the CKE to be:
\begin{equation}\begin{split}
&	\Big[(u\cdot p) \mp \hbar \frac{ \omega \cdot p}{2\, u \cdot p}\Big] \hat{\mathrm{d}} f_\pm
	 - \mathcal{C}_\pm[f_+, f_-] \\
=\,&	- p^\mu \nabla_\mu f_\pm
	\mp \hbar \Big(\frac{\epsilon^{\mu\nu\rho\sigma} p_\nu (\partial_\rho  u_\sigma)}{4\, u \cdot p}\Big)\nabla_\mu f_\pm  \,,
\end{split}\label{eq.f_eom_2}
\end{equation}
where $\hat{\mathrm{d}} X \equiv u^\mu \partial_\mu X$, $\nabla_\mu \equiv \Delta_{\mu\nu} \partial^\nu$.
In the 14-moment expansion formalism, we expand the non-equilibrium correction to be moments of $p_{\langle\alpha}\cdots p_{\beta\rangle}$, and truncate terms up to $p^2$ order:
\begin{equation}
\begin{split}
f^\pm \equiv\,& f_\mathrm{eq}^\pm + f_\mathrm{eq}^\pm (1-f_\mathrm{eq}^\pm) 
	\bigg[ 
\\&\quad	+	\lambda_{\Pi}^\pm \Pi
		+	\lambda_{\nu}^\pm \nu_{\pm}^\mu p_\mu
		+	\lambda_{\pi}^\pm \pi^{\mu\nu} p_\mu p_\nu 
	\bigg] \label{eq.non_equilibrium_distribution}\\
=\,& f_0^\pm + f_0^\pm (1-f_0^\pm)
	\bigg[ 
		\mp	\frac{\hbar}{2T}\frac{\omega \cdot p}{u\cdot p}
\\&\quad	+	\lambda_{\Pi}^\pm \Pi
		+	\lambda_{\nu}^\pm \nu_{\pm}^\mu p_\mu 
		+	\lambda_{\pi}^\pm \pi^{\mu\nu} p_\mu p_\nu 
\bigg] \,, 
\end{split}
\end{equation}
where
\begin{eqnarray}
f_{0,\pm}(p) &=&  \frac{1}{\exp[\frac{u \cdot p - \mu_\pm}{T}]+1}\,,\\
f_{\mathrm{eq},\pm}(p) 
&=&	\frac{1}{\exp[\frac{u \cdot p - \mu_\pm}{T} \pm \hbar\frac{1}{2T} \frac{\omega \cdot p}{u\cdot p}]+1} \nonumber\\
&=&	 f_{0,\pm} \mp f_{0,\pm}(1-f_{0,\pm})\frac{\hbar}{2T}\frac{\omega \cdot p}{u\cdot p} + \mathcal{O}(\hbar^2) .\quad\;
\end{eqnarray}
Noting that the equilibrium form of polarization vector $\omega^\mu$ is a first-order derivative term, we keep up to first order in viscous expansion. This is consistent with the order of quantum corrections.


It is worth noting that in the above expressions, $T$ and $\mu_\pm$ are the effective temperature and chemical potentials, respectively.
In principle, these quantities are well defined only in thermal systems; whereas in practice, one can define them for non-equilibrated systems by matching the energy and particle densities
\begin{eqnarray}
\epsilon &\equiv& \int_p (u\cdot p)^2 [f_{+}(p) + f_{-}(p)]  \,,\label{eq.definition_epsilon}\\
n_\pm &\equiv&  \int_p (u\cdot p) f_\pm(p)  \,,\label{eq.definition_n}
\end{eqnarray}
with their corresponding equilibrium expectations:
\begin{eqnarray}
\epsilon &=& \epsilon_\mathrm{eq}  \equiv \int_p (u\cdot p)^2 [f_{\mathrm{eq},+}(p)+f_{\mathrm{eq},-}(p)] \,,\label{eq.match_epsilon}\\
n_\pm &=& n_\mathrm{eq,\pm}  \equiv \int_p (u\cdot p) f_{\mathrm{eq},\pm}(p)\,.
\end{eqnarray}

With these, one can separate the pressure into two parts --- the thermal pressure $P$, and the bulk pressure $\Pi$ being the non-equilibrium correction:
\begin{eqnarray}
P &\equiv& - \frac{1}{3}\int_p \Delta^{\mu\nu} p_\mu p_\nu [f_{\mathrm{eq},+}(p)+f_{\mathrm{eq},-}(p)] \,,\label{eq.definition_P}\\
\Pi  &\equiv& - \frac{1}{3}\int_p \Delta^{\mu\nu} p_\mu p_\nu [\delta f_{+}(p) + \delta f_{-}(p)] \,,\qquad\label{eq.definition_Pi}
\end{eqnarray}
where $\delta f_\pm \equiv f^\pm - f_\mathrm{eq}^\pm$ denotes the non-equilibrium sector of the distribution functions.
Implementing the energy matching relation \eqref{eq.match_epsilon}, one can re-express Eq.\eqref{eq.definition_Pi} as
\begin{equation}\begin{split}
\Pi =\,& - \frac{1}{3}\int_p (p^\mu p_\mu)  [\delta f_{+}(p) + \delta f_{-}(p)] \\
=\,& -\frac{m^2}{3} \int_p [\delta f_{+}(p) + \delta f_{-}(p)]\,.
\end{split}\end{equation}
In the massless limit $m^2=0$, the bulk viscous pressure vanishes, hence the scalar corrections $\lambda_\Pi^\pm \Pi$ disappear.

Besides, one can further define the non-equilibrium corrections to hydrodynamics --- the dissipative quantities:
\begin{eqnarray}
\pi^{\mu\nu} &\equiv& \int_p \Delta^{\mu\nu}_{\alpha\beta} p^\alpha p^\beta [f_{+}(p) + f_{-}(p)] \,, \label{eq.definition_pi}\\
\nu_\pm^\mu &\equiv& \int_p \Delta^\mu_\alpha p^\alpha \delta f_\pm (p) \,.\label{eq.definition_nu}
\end{eqnarray}
From the relations in Eq.(\ref{eq.definition_epsilon} - \ref{eq.definition_nu}), one can fix the coefficients in non-equilibrium distribution function:
\begin{equation}\begin{split}
&	\lambda_{\pi}^\pm = \frac{1}{4J_{4,2}^\pm} \,,\qquad
	\lambda_{\nu}^\pm =  \frac{J_{3,1}^\pm (u \cdot p) - J_{4,1}^\pm}{D_{3,1}^\pm} \,.
\end{split}\end{equation}
Detailed derivations can be found in Appendix~\ref{appendix.moment_coefficients}.

Substituting the distribution function in the definition~(\ref{eq.current}-\ref{eq.stresstensor}), we find the RH and LH particle currents and energy-momentum stress tensor:
\begin{eqnarray}
J_{\pm}^{\mu} 
&=&n_\pm u^\mu + \nu_\pm^\mu 
	\pm \frac{\hbar}{2} \frac{\partial n_\pm}{\partial \mu_\pm} \omega^\mu
\nonumber\\&&
	\pm \frac{\hbar}{2} \epsilon^{\mu\rho\sigma\lambda} u_\rho \partial_\sigma 
		\Big(\frac{G_{4,1}^{(1),\pm}}{D_{3,1}^\pm} \nu_{\pm,\lambda}\Big)
\nonumber\\&&
	\pm \frac{\hbar J_{2,2}^\pm}{4J_{4,2}^\pm} \Big( \epsilon^{\mu\rho\sigma\lambda} 
		u_\rho {\sigma_{\sigma}}^{\xi}\pi_{\lambda\xi} - \pi^{\mu\lambda} \omega_{\lambda} \Big)
\nonumber\\ &\equiv& 
	n_\pm \, u^\mu + \nu_\pm^\mu + \hbar\, J_\mathrm{quantum,\pm}^\mu
	\,,
\label{eq.2nd_current}
\end{eqnarray}

\begin{eqnarray}
T^{\mu\nu} 
&=&	\varepsilon \, u^\mu u^\nu - P \,\Delta^{\mu\nu} + \pi^{\mu\nu} 
	+ \frac{4\hbar}{5}\omega^\mu (\nu_+^{\nu} - \nu_-^{\nu}) 
\nonumber\\&&
	+\frac{\hbar \,n_A}{4} (8 \omega^\mu u^\nu + T \epsilon^{\mu\nu\sigma\lambda}\varpi_{\sigma\lambda})
\nonumber\\&&
	+ \frac{\hbar}{2} \epsilon^{\mu\rho\sigma\lambda} u_\rho \Delta^{\nu\xi}\partial_\sigma\bigg[
		 \Big(\frac{J_{3,2}^+}{2J_{4,2}^+} -\frac{J_{3,2}^-}{2J_{4,2}^-} \Big)\pi_{\lambda\xi}  \bigg]
\nonumber\\&&
	+  \frac{\hbar}{2} \epsilon^{\mu\rho\sigma\lambda} u_\rho u^\nu
		 \partial_\sigma(\nu^+_{\lambda} -  \nu^-_{\lambda})
\nonumber\\&&
	- \frac{\hbar}{10} \epsilon^{\mu\nu\rho\sigma} u_\rho 
		(\partial_\sigma u^\lambda) (\nu^+_{\lambda} - \nu^-_{\lambda})
\nonumber\\&&
	+ \frac{2\hbar}{5} \epsilon^{\mu\lambda\rho\sigma} u_\rho 
		(\partial_\sigma u^\nu) (\nu^+_{\lambda} - \nu^-_{\lambda})
\nonumber\\ &\equiv& 
	\varepsilon \, u^\mu u^\nu - P\,\Delta^{\mu\nu} + \pi^{\mu\nu} + \hbar \, T_\mathrm{quantum}^{\mu\nu} 
	 \,.
\label{eq.2nd_stress}
\end{eqnarray}
Together with classical dissipation terms $\pi^{\mu\nu}$ and $\nu_\pm^\mu$, viscous corrections also modify the quantum  $T_\mathrm{quantum}^{\mu\nu}$ and $ J_\mathrm{quantum,\pm}^\mu$, from their equilibrium form.
In this work, we take the Landau frame and define flow velocity $u^\mu$ as the time-like left-eigenvector of the stress tensor, with energy density $\epsilon$ being the eigenvalue:
\begin{equation}
u_\mu T^{\mu\nu}_\mathrm{classical}  = \epsilon \, u^\nu \,.
\end{equation}

Finally, we derive the equations of motion for dissipative terms, ruled by:
\begin{eqnarray}
\Delta^{\mu\nu}_{\rho\sigma} \hat{\mathrm{d}}\pi^{\rho\sigma} &\equiv& 
	\int_p \Delta^{\mu\nu}_{\alpha\beta} p^\alpha p^\beta \Big( \hat{\mathrm{d}} \delta f_{+} + \hat{\mathrm{d}} \delta f_{-}\Big) \,, \\
\Delta^{\mu\nu} \hat{\mathrm{d}} \nu_{\pm,\nu} &\equiv& \int_p \Delta^\mu_\alpha p^\alpha \hat{\mathrm{d}}\delta  f_\pm \,,
\end{eqnarray}
whereas the equation of motion for $\delta f_\pm$ is derived from Eq.~\eqref{eq.f_eom_2}:
\begin{equation}\begin{split}
& \hat{\mathrm{d}} \delta f_\pm 
	- \Big( \frac{1}{u\cdot p} \pm \hbar \frac{\omega\cdot p}{2\, (u \cdot p)^3} \Big) \mathcal{C}_\pm[f_+, f_-] 
\\ =&
	- \hat{\mathrm{d}} f_{\mathrm{eq},\pm} 
 	- \frac{p^\mu \nabla_\mu f_\pm}{u\cdot p}
\\ &
	\mp \frac{\hbar  \epsilon^{\mu \nu \lambda \sigma} p_\nu p^\rho u_\lambda (\partial_\rho u_\sigma - \partial_\sigma u_\rho)}{4(u \cdot p)^3} \nabla_\mu f_\pm\,.
\end{split}\end{equation}

Putting the lengthy calculations in App.~\ref{app.eom_dissipative} and keeping up to second-order terms, the relaxation equations for all the dissipative terms are
\begin{eqnarray}
&&\begin{split}
&	\Delta^{\alpha\beta}_{\rho\sigma} \hat{\mathrm{d}}\pi^{\rho\sigma} 
	- (\mathcal{A}_{+,0}^{(2)} + \mathcal{A}_{-,0}^{(2)}) \pi^{\alpha\beta}	 
\\&\quad
	- \frac{\hbar}{2} (\mathcal{X}_{2,-2}^{+,+} - \mathcal{X}_{2,-2}^{-,+}) 
		\Delta^{\alpha\beta}_{\rho\sigma} \omega^\rho \nu_+^\sigma
\\&\quad
	+ \frac{\hbar}{2} (\mathcal{X}_{2,-2}^{-,-} -  \mathcal{X}_{2,-2}^{+,-}) 
		\Delta^{\alpha\beta}_{\rho\sigma} \omega^\rho \nu_-^\sigma 
\\=\,& 
	\frac{8}{5} P \sigma^{\alpha\beta}
	- 3\theta\,\pi^{\alpha\beta} 
	+ \frac{8}{7} \Delta^{\alpha\beta} \sigma^{\mu\nu} \pi_{\mu\nu} 
	-\frac{12}{7} \sigma^{\alpha}_{\;\;\mu} \pi^{\beta\mu}
\\&
	-\frac{12}{7} \sigma^{\beta}_{\;\;\mu} \pi^{\alpha\mu}
	-\pi^{\alpha}_{\;\;\mu}\epsilon^{\beta\mu\nu\rho} u_\nu \omega_\rho
	- \pi^{\beta}_{\;\;\mu}\epsilon^{\alpha\mu\nu\rho} u_\nu \omega_\rho
\\&
	+ \frac{2\hbar}{15} \Delta^{\alpha\beta}_{\mu\nu} \omega^\mu \nabla^\nu n_A 
	+ \frac{\hbar}{5}  n_A \,\Delta^{\alpha\beta}_{\mu\nu} \nabla^{\mu} \omega^\nu 
\\&
	- \frac{9\hbar}{10}\frac{n_A}{\varepsilon+P} \Delta^{\alpha\beta}_{\mu\nu} \omega^\mu \nabla^\nu P
\\&
	+ \frac{\hbar}{20}\frac{n_A}{\varepsilon+P} \Big(\sigma_{\;\;\mu}^{\beta} 
		\epsilon^{\mu \alpha \lambda \sigma} u_\lambda \nabla_\sigma P
	+\sigma_{\;\;\mu}^{\alpha} \epsilon^{\mu \beta \lambda \sigma} u_\lambda \nabla_\sigma P \Big)\,,
\end{split}\label{eq.2nd_pi}
\end{eqnarray}
and
\begin{eqnarray}
&&\begin{split}
&	\Delta^{\alpha\beta} \hat{\mathrm{d}}\nu_{\beta} ^{\pm}
	- \mathcal{A}_{\pm,0}^{(1)} \nu^\alpha_\pm
	- \mathcal{B}_{\pm,0}^{(1)} \nu^\alpha_\mp
	\pm \frac{\hbar}{2T} \mathcal{W}_{\pm,0}^{(1)} \omega^\alpha
\\&\quad
	+ \frac{\hbar}{2} \Big(\mathcal{A}_{+,-2}^{(2)} - \mathcal{A}_{-,-2}^{(2)}\Big) \pi^{\alpha\beta}\omega_\alpha
\\=\,&
	\frac{D_{2,1}^\pm}{J_{3,1}^\pm}\nabla^\alpha \frac{\mu_\pm}{T}
	+\ \frac{D_{3,0}^\pm}{2J_{3,0}^\pm J_{4,0}^\pm} \Delta^{\alpha}_{\rho} \nabla_\mu \pi^{\mu\rho}
	-\pi^{\alpha\mu} \nabla_\mu  \frac{J_{3,0}^\pm}{2J_{4,0}^\pm}
\\&
	- \theta \nu_\pm^\alpha - \frac{3}{5}\sigma^{\alpha\mu} \nu^\pm_\mu 
	- \epsilon^{\alpha\mu\nu\gamma} u^\mu \nu_\pm^\nu \omega^\gamma 
\\&
	\mp  \frac{\hbar}{3} \omega^\alpha \hat{\mathrm{d}} I_{0,0}^{\pm}
	\mp \frac{\hbar}{2T} \frac{D_{2,1}^\pm}{J_{3,1}^\pm} \Delta^\alpha_{\beta} \hat{\mathrm{d}} \omega^\beta
\\&
	\pm \frac{3\hbar}{2} \frac{n_\pm^2}{\varepsilon_\pm+P_\pm}
		 \Big( \frac{1}{3}\theta \omega^\alpha + \sigma^{\alpha\mu} \omega_\mu \Big)
\\&
	\mp\frac{\hbar}{3} I_{0,0}^\pm \Big( \frac{13}{15}\theta \omega^\alpha + \frac{4}{5}\sigma^{\alpha\mu} \omega_\mu \Big)
\\&
	\pm \frac{\hbar}{12} \epsilon^{\mu \alpha \lambda \sigma} u_\lambda \hat{\mathrm{d}} u_\sigma (\nabla_\mu I_{0,0}^\pm)\,,
\end{split}\label{eq.2nd_nu}
\end{eqnarray}
where $\mathcal{A}, \mathcal{B}, \mathcal{W}, \mathcal{X}$ are integrals of collision kernel defined in App.~\ref{app.collision_kernels}. 
They are functions of temperature $T$ and chemical potentials $\mu_\pm$.
We note that there have been similar attempts to derive the dissipative spin hydrodynamics from the relaxation-time approximation~\cite{Yang:2018lew,Bhadury:2020puc}, i.e. the collision kernel is approximated by $(f-f_\mathrm{eq})/\tau_\mathrm{eq}$.
We emphasize that by taking the 14-moment formalism with a concrete collision kernel, we are able to obtain the exact form of transport coefficients and relaxation times. 
In this paper, we aim to construct a theoretical framework based on the general form of collision terms.
Recent studies --- focusing on relativistic heavy-ion collisions --- of the relaxation time can be found in~\cite{Kapusta:2019sad,Ayala:2019iin,Ayala:2020ndx}.

We end by discussing the viscous correction to the spin degrees of freedom.
At the macroscopic level, the spin density at the fluid co-moving frame is
\begin{eqnarray}
\begin{split}
S^{\mu\nu} \equiv&\; u_\lambda S^{\lambda\mu\nu} \\
 =&\;	\frac{1}{2} \epsilon^{\sigma\lambda\mu\nu}\int_p u_\lambda \mathcal{A}_\sigma \\
=&\;	
	\frac{1}{2} \epsilon^{\sigma\lambda\mu\nu} u_\lambda (J_{+,\sigma} - J_{-,\sigma}) \\
=&\; 
	\frac{\hbar \, T}{4}\Big(\frac{\partial n_+}{\partial \mu_+} + \frac{\partial n_-}{\partial \mu_-}\Big)
		\Delta^{\mu}_{\alpha} \Delta^{\nu}_{\beta} \varpi^{\alpha\beta}
	+ \frac{1}{2} \epsilon^{\mu\nu\sigma\lambda} \nu_{A,\sigma} u_\lambda
\\&
	+ \frac{\hbar}{2} \epsilon^{\mu\nu\sigma\lambda} \epsilon_{\sigma\alpha\beta\gamma} u_\lambda u^\alpha \partial^\beta 
		\bigg(\frac{G_{4,1}^{(1),+}}{D_{3,1}^+} \nu_{+}^{\gamma} + \frac{G_{4,1}^{(1),-}}{D_{3,1}^-} \nu_{-}^{\gamma}\bigg)
\\&
	+ \frac{\hbar}{4}\Big( \frac{J_{2,2}^+}{J_{4,2}^+} + \frac{J_{2,2}^-}{J_{4,2}^-} \Big)
	(\pi^{\mu\xi}  {\sigma^{\nu}}_{\xi} - \pi^{\nu\xi}  {\sigma^{\mu}}_{\xi})
\\&	- \frac{\hbar}{4}\Big( \frac{J_{2,2}^+}{J_{4,2}^+} + \frac{J_{2,2}^-}{J_{4,2}^-} \Big)
	\epsilon^{\mu\nu\sigma\lambda} u_\lambda \pi_{\sigma\alpha} \omega^{\alpha}\,.
\end{split}
\end{eqnarray}
Especially, in the equilibrium limit that all viscous corrections are turned-off, i.e. $\nu^\mu\to0$, $\pi^{\mu\nu}\to0$, the spin density $S^{\mu\nu} \propto \Delta^{\mu}_{\alpha} \Delta^{\nu}_{\beta} \varpi^{\alpha\beta}$ is proportional to the spatial components of thermal vorticity tensor.

At the microscopic level, one would be interested in the polarization rate for individual particles, especially for final hadrons.
The momentum-dependent mean spin vector for each hadron can be obtained as follows (see e.g. \cite{Becattini:2020sww}),
\begin{eqnarray}\begin{split}
S^{\mu}(p) 
=&\; -\frac{1}{8}\epsilon^{\mu\nu\rho\sigma} p_\nu 
	\frac{\int \mathrm{d}\Sigma_{\mathrm{fo},\lambda}  \mathrm{tr}[ \{\gamma^\lambda, \Sigma_{\rho\sigma} \}W(x,p)]}
	{\int \mathrm{d}\Sigma_{\mathrm{fo},\lambda} p^\lambda \mathrm{tr}[W(x,p)]} \\
=&\; \frac{1}{4m_H}\epsilon^{\mu\nu\rho\sigma} p_\nu 
	\frac{\int \mathrm{d}\Sigma_{\mathrm{fo}}^{\lambda}  \epsilon_{\lambda\rho\sigma\delta}  \mathcal{A}^\delta (x,p)}
	{\int \mathrm{d}\Sigma_{\mathrm{fo},\lambda} \mathcal{V}^\lambda(x,p)} \\
=&\; \frac{1}{2m_H}
	\frac{\int \mathrm{d}\Sigma_{\mathrm{fo}}^{\lambda} p_\lambda \mathcal{A}^\mu (x,p)}
	{\int \mathrm{d}\Sigma_{\mathrm{fo}}^{\lambda} \mathcal{V}_\lambda(x,p)} \,,
\end{split}\end{eqnarray}
where $\Sigma_{\mathrm{fo},\lambda}$ represents the freeze-out hyper-surface.
Assuming that hadrons take the same distribution as the 14-moment formalism~(\ref{eq.non_equilibrium_distribution}), we find
\begin{widetext}
\begin{eqnarray}\begin{split}
S^{\mu}(p) 
=&\;	 \frac{1}{2m_H}
	\bigg\{ \Big[\int_\Sigma f_{V,0}\Big] +
	 \int_\Sigma f_{V,0}(1-f_{V,0})(\lambda_{\nu} \nu^\alpha p_\alpha 
	+\lambda_{\pi} \pi^{\alpha\beta} p_\alpha p_\beta ) \bigg\}^{-1}
\\&\;\times\bigg\{
	\Big[ - \frac{\hbar}{4} \epsilon^{\mu\nu\rho\sigma} \int_\Sigma p_\nu \varpi_{\rho \sigma} f_{V,0} (1-f_{V,0}) \Big]
	+ \int_\Sigma p^\mu f_{V,0}(1-f_{V,0})  \frac{\mu_A}{T}
\\&\;\quad	+ \int_\Sigma p^\mu f_{V,0}(1-f_{V,0}) 
		\Big(\frac{\lambda_{\nu}}{2} \nu_A^\alpha p_\alpha 
		  + \frac{\lambda_{\nu}^+ - \lambda_{\nu}^-}{2} \nu^\alpha p_\alpha 
	+\frac{\lambda_{\pi}^+ - \lambda_{\pi}^-}{2} \pi^{\alpha\beta} p_\alpha p_\beta 
	\Big) \bigg\} + \mathcal{O} (\hbar^2) \,,
\end{split}\label{eq.spin_vector}\end{eqnarray}
\end{widetext}
where $f_{V,0} \equiv [e^{(u \cdot p -\mu)/T} + 1]^{-1}$ is the Fermi-Dirac distribution, $\int_\Sigma (\cdots) \equiv \int_\Sigma \mathrm{d}\Sigma_{\mathrm{fo}}^{\lambda} p_\lambda (\cdots)$ is the integral over freeze-out hyper-surface,
and
\begin{eqnarray}\begin{split}
	\mu_{A} \equiv (\mu_{+} - \mu_{-})/2\,,&\qquad
	\mu \equiv (\mu_{+} + \mu_{-})/2\,,\\
	\nu_{A}^\mu \equiv \nu_{+}^\mu - \nu_{-}^\mu\,,&\qquad
	\nu^\mu \equiv \nu_{+}^\mu + \nu_{-}^\mu\,.
\end{split}
\end{eqnarray}
In the expression of the mean spin vector per particle (\ref{eq.spin_vector}), if keeping terms in $[ \cdots ]$ only, one can repeat the equilibrium result in Ref.~\cite{Becattini:2020sww}, whereas the other terms are corrections.
Among them, there is a term proportional to $\mu_A /T$, which is a leading order contribution, in both gradient expansion and semi-classical expansion.
It acts oppositely for $\Lambda$ and $\bar\Lambda$ hyperons, and 
might suggest an explanation to the measured difference in their polarization rate~\cite{STAR:2017ckg}.
The rest of the terms are viscous corrections: The ones in the denominator, $\{\cdots\}^{-1}$, are corrections to spin-averaged particle distribution; whereas the ones in the numerator are corrections directly to the spin distribution.
The latter might be related to the sign difference between theory and experiment results on azimuthal angle distribution of longitudinal polarization.
Last but not least, noting that for systems starting with zero chirality imbalance, all quantities proportional to the difference between right and left, i.e. $\mu_A$ and $\nu_A^\mu$, appear because of chiral transport, hence are proportional to $\hbar$. 
Therefore, such terms are consistent in both quantum and viscous expansions.
\section{Summary and Outlook}


In this work, we start from a 14-moment expansion formalism and obtain the second-order viscous spin hydrodynamics from a system of massless Dirac spinors.
In such a system, the spin alignment effect could be treated in the same framework as for chiral hydrodynamics, but with non-trivial quantum corrections to the stress tensor.
We further obtain the non-equilibrium correction to the spin polarization vector, and find a potential new source for the difference in the polarization rate of $\Lambda$ and $\bar\Lambda$ hyperons.

We construct a hydrodynamic theory that self-consistently solves the evolution of systems containing spin degrees of freedom and includes the viscous-corrections in the hadron spin polarization rate, and the explicit form of the hydrodynamics quantities and equations are shown in Eqs.~(\ref{eq.2nd_current}, \ref{eq.2nd_stress}, \ref{eq.2nd_pi}, \ref{eq.2nd_nu}).
This framework will be implemented in future numerical hydrodynamic simulations to precisely quantify both global and local polarization rates of final-state hadrons created in heavy-ion collisions.

We need to point out that whereas taking the chiral limit, both the spin tensor and the axial current can be represented by the semi-conserved axial charge. For massive fermions, on the other hand, one would need to introduce another two independent degrees of freedom to construct the microscopic state~\cite{Weickgenannt:2019dks,Hattori:2019ahi,Guo:2020zpa,Sheng:2020oqs}. To fully explore the spin dynamics for a generic system, one would need to start from the quantum kinetic theory for massive particles to construct the corresponding viscous hydrodynamic theory. This would be performed in our future work.

We end by noting that hydrodynamic theory is a macroscopic theory that can be derived from conservation laws and the second law of thermodynamics. 
A hydrodynamic theory containing the spin degrees of freedom has been constructed based on such macroscopic principles in Ref.~\cite{Hattori:2019lfp}.
It is particularly interesting to compare the results derived from a microscopic approach to those derived from a macroscopic approach.
Compared to the results of~\cite{Hattori:2019lfp} where parity-odd effects are not considered, we find extra terms could be added without violating conservation laws and entropy production law.
Those results will be reported in a separate publication.

\textbf{Acknowledgments --- }
This work was supported by the Natural Sciences and Engineering Research Council of Canada.
S.S. also acknowledges support from the Fonds de Recherche du Qu\'ebec - Nature et Technologies (FRQNT) through the Programmede Bourses d'Excellencepour \'Etudiants \'Etrangers (PBEEE). We would like to thank Francesco Becattini, Wojciech Florkowski, Xu-Guang Huang, Jinfeng Liao, Jorge Noronha, Dirk Rischke, Qun Wang, and Yi Yin for helpful discussions.

\begin{appendix}
\section{Stability and Causality of Spin Fluid Dynamics}\label{sec.sound_mode}

A unique feature of spin hydrodynamics is the emergence of vorticity vector $\omega^\mu$ terms at ideal order, which is a first-order derivative of velocity $u^\mu$.
Given this, one may be concerned by the numerical stability and relativistic causality of the theory.
Generally speaking this is not an issue, as the definition of vorticity vector contains anti-symmetric Levi-Civita tensor, hence neither $\partial_\mu \omega^\mu$ nor $\omega^\mu \partial_\mu X$ contain second-order derivative terms, not even the product of first order terms with respect to the same variable.
To see this, we follow the procedure in Ref.~\cite{Hiscock:1985zz} and examine the linear perturbation on top of a homogenous-constant background.
Without loss of generality, we take the direction of the background fluid velocity as the $\hat{z}$-direction, hence the full velocity is $u^\mu = \gamma(1,0,0,\beta) + (\delta u^t,\delta u^x,\delta u^y,\delta u^z)$, with $\gamma\equiv(1-\beta^2)^{-1/2}$ being the Lorentz factor.
Similarly, the full energy density becomes $\varepsilon+\delta\varepsilon$, while number density is $n_V + \delta n_V$, axial number density is $n_A + \delta n_A$.
Noting that the four-velocity must be normalized, $u_\mu u^\mu = 1$, hence $\delta u^t - \beta \delta u^z = 0$. It would be more convenient to let $\delta u^z = \gamma \delta u^3$ and $\delta u^t = \gamma \beta \delta u^3$, and we label $\delta u^x = \delta u^1$ and $\delta u^y = \delta u^2$ for consistency. One can see that $\delta u^1$, $\delta u^2$, and $\delta u^3$ correspond to $\delta {\bf u}$ in the fluid co-moving frame.

The evolution of the perturbative quantities, $\{\delta \varepsilon, \delta n_V, \delta n_A, \delta u^1, \delta u^2, \delta u^3 \}$, is governed by:
\begin{equation}
\partial_\mu \delta J_{V}^{\mu}  = 0,\qquad
\partial_\mu \delta J_{A}^{\mu}  = 0,\qquad
\partial_\mu\delta T^{\mu\nu} = 0.
\end{equation}
Expanding the hydrodynamic equations for linear perturbations, one finds:
\begin{align}
\begin{split}
0 =& 
	\gamma (\partial_t + \beta \partial_z) \delta n_V
\\&	+ n_V (\partial_x \delta u^1 + \partial_y \delta u^2 
	+  \gamma( \beta \partial_t  + \partial_z) \delta u^3) \,,
\end{split}\label{eq.sound1}\\
\begin{split}
0 =& 
	\gamma (\partial_t + \beta \partial_z) \delta n_A
\\&	+ n_A (\partial_x \delta u^1 + \partial_y \delta u^2 
	+  \gamma( \beta \partial_t  + \partial_z) \delta u^3) \,,
\end{split}\\
\begin{split}
0 =& 
	\gamma (\partial_t + \beta \partial_z) \delta \epsilon
\\&	+ H (\partial_x \delta u^1 + \partial_y \delta u^2 
	+  \gamma(\beta \partial_t  + \partial_z) \delta u^3) \,,
\end{split}\\
\begin{split}
0 =& 
	H \gamma(\partial_t + \beta \partial_z) \delta u^1 + \partial_x\delta P 
\\&	+ \frac{\hbar \, n_A }{2} \gamma(\partial_t + \beta \partial_z) 
	(\partial_y \delta u^3 - \gamma(\beta \partial_t + \partial_z) \delta u^2) \,,
\end{split}\\
\begin{split}
0 =& 
	H \gamma(\partial_t + \beta \partial_z) \delta u^2 + \partial_y\delta P 
\\&	+ \frac{\hbar \, n_A }{2} \gamma(\partial_t + \beta \partial_z) 
	(\gamma(\beta \partial_t + \partial_z) \delta u^1 - \partial_x \delta u^3) \,,
\end{split}\\
\begin{split}
0 =& 
	H  \gamma^2(\partial_t + \beta \partial_z) \delta u^3 
	+  \gamma^2(\beta \partial_t + \partial_z)\delta P 
\\&	+ \frac{\hbar \, n_A }{2} \gamma^2(\partial_t + \beta \partial_z) 
	 (\partial_x \delta u^2 - \partial_y \delta u^1) \,.
\end{split}\label{eq.sound6}
\end{align}
where $H \equiv \varepsilon + P$ is the enthalpy.
Compared to the ``spinless'' hydro, the evolution equations contain second-order derivative terms $(\hbar \, n_A/2)\partial_\mu \partial_\nu \delta u^\rho$.
However, this does not necessarily mean instability or acausality.
To see it explicitly, we apply Fourier transformation to the perturbative quantities and solve the plane-wave eigenmodes:
\begin{eqnarray}
\left[\begin{array}{c}
\delta \varepsilon \\
\delta n_V \\
\delta n_A \\
\delta u^1 \\
\delta u^2 \\
\delta u^3 \\
\end{array}\right] = \exp[i (\omega t  - k_x x - k_y y - k_z z)]
\left[\begin{array}{c}
\delta \varepsilon_0 \\
\delta n_{V0} \\
\delta n_{A0} \\
\delta u^1_0 \\
\delta u^2_0 \\
\delta u^3_0 \\
\end{array}\right] .\quad \label{eq.plane_wave}
\end{eqnarray}
For later convenience, we apply the variable substitution $\omega=\gamma(\omega' + \beta k_z')$ and $k_z = \gamma(\beta\omega' + k_z')$. Then the plane-wave becomes
\begin{align}\begin{split}
&\; \exp[i (\omega t -  k_x x - k_y y -k_z z)] \\
 =&\; \exp[ i [\omega' \gamma(t - \beta z) - k_x x - k_y y - k_z' \gamma(z-\beta t)]],
\end{split}\end{align}
and $k_z'$ and $\omega'$ respectively correspond to the wavenumber in $z$-direction and frequency in the fluid co-moving frame.
For the plane-wave modes, one can make the replacement
\begin{align}
&\partial_x \to -i\, k_x,\qquad
\partial_y \to -i\, k_y,\qquad\\
&\partial_t \to i\, \gamma(\omega' + \beta k_z'),\qquad
\partial_z \to -i\, \gamma(\beta\omega' + k_z'),\\
&\gamma (\partial_t + \beta \partial_z) \to i\, \omega' ,\qquad
\gamma(\beta \partial_t + \partial_z) \to - i\, k_z',
\end{align}
in the equations (\ref{eq.sound1} - \ref{eq.sound6}), and rewrite them as
\begin{equation}
\left[\begin{array}{cccccc}
- \omega' & 0 & 0 & a\,k_x & a\,k_y & a\,k_z' \\
0 & - \omega' & 0 & b\,k_x & b\,k_y & b\,k_z' \\
0 & 0 & - \omega' & c\,k_x & c\,k_y & c\,k_z' \\
d\,k_x & e\,k_x & f\,k_x & - \omega' & - g^*\omega'\,k_z' & - g\,\omega'\,k_y \\ 
d\,k_y & e\,k_y & f\,k_y & - g\,\omega'\,k_z' & - \omega' & - g^*\omega'\,k_x \\ 
d\,k_z' & e\,k_z' & f\,k_z' & - g^*\omega'\,k_y & - g\,\omega'\,k_x & - \omega'  \\ 
\end{array}\right] \cdot
\left[\begin{array}{c}
\delta \varepsilon \\
\delta n_V \\
\delta n_A \\
\delta u^1 \\
\delta u^2 \\
\delta u^3 \\
\end{array}\right] = 0\;,\label{eq.matrix}
\end{equation}
where
\begin{equation}\begin{split}
&a \equiv \varepsilon + P, \quad
  b \equiv n_V,\quad
  c \equiv n_A,\quad
  d \equiv \frac{1}{H} \frac{\partial P}{\partial \varepsilon} ,\\
&e \equiv \frac{1}{H} \frac{\partial P}{\partial n_V} ,\qquad
   f \equiv \frac{1}{H} \frac{\partial P}{\partial n_A} ,\qquad
   g \equiv i \cdot \frac{\hbar \, n_A }{2H} .
\end{split}\end{equation}
Particularly, $g$ is purely imaginary, and $g^*=-g$. The six eigenvalues of the coefficient matrix~\eqref{eq.matrix} are:
\begin{equation}
\omega',\quad \omega',\quad 
\omega' \pm |g|\,k' \,\omega',\quad
\omega' \pm \sqrt{ad + be + cf}k',\label{eq.eigenvalues}
\end{equation}
with $k'\equiv\sqrt{k_x^2 + k_y^2 + k_z'^2}$.
The solution of perturbation field would be trivial unless one of the above eigenvalues is zero.
Such a condition leads to the constraint-equation between $\omega$ and ${\bf k}$ --- the latter is also referred to as the dispersion relation. 
For the eigenvalues in \eqref{eq.eigenvalues}, we note that $|g|\,k' = \hbar\,n_A k'/(2H) \ll 1$ per the requirement of semi-classical expansion, hence $1\pm |g|\,k' \neq0$, and $\omega' \pm |g|\,k'\,\omega' = 0$ leads to $\omega'=0$.
With these, non-trivial modes can be found if
\begin{equation}
\omega' = 0,\qquad \text{or}\quad\omega'=\pm c_s k'. \label{eq.constraint}
\end{equation}
Particularly, the speed of sound in the fluid co-moving frame,
\begin{equation}\begin{split}
c_s \equiv&\; \sqrt{ad + be + cf} \\
=&\; \Big(\frac{\partial P}{\partial \varepsilon} + \frac{n_V}{\varepsilon + P} \frac{\partial P}{\partial n_V} + \frac{n_A}{\varepsilon + P} \frac{\partial P}{\partial n_A} \Big)^{1/2},
\end{split}\end{equation}
is determined by the equation of state and takes the same formula as the ``spin-less'' hydro.
Re-expressing the constraint equations~\eqref{eq.constraint} with lab-frame quantities, the dispersion relations of the non-vanishing modes are:
\begin{equation}
\omega = \beta k_z, \label{eq.static}
\end{equation}
or
\begin{equation}
\omega = \frac{ (1-c_s^2)\beta k_z \pm c_s \gamma^{-2}\sqrt{k_z^2 + (1-\beta^2c_s^2)\gamma^2 k_\perp^2} }{1-\beta^2c_s^2}. \label{eq.sound_doppler}
\end{equation}
It is clear that \eqref{eq.static} is the ``static'' perturbation moving together with the fluid background, while \eqref{eq.sound_doppler} is the sound-propagation with Doppler effect.
The property of Lorentz transformation ensures the speed of sound to be less than the speed of light.
Consequently, one can conclude that spin hydrodynamics equations remain causal and is stable for linear perturbations, even though they contain the derivative term $\omega^\mu$.

We end this section by noting that, in general, the causality and stability of linearized sound modes do not guarantee the causality and stability of the whole theory --- far-from-equilibrium perturbations can not be approximated as linearized modes. Therefore, our study can be considered a necessary, but non-sufficient condition for stability. A complete analysis takes into account the non-linear far-from-equilibrium perturbations. One then may need the techniques recently developed in Ref.~\cite{Bemfica:2020xym}. This lies beyond the scope of this project, and is left for future work. 

\section{Pseudo-Gauge Transformation to Symmetrize The Energy-Momentum Tensor} \label{sec.pseudogauge}

It is worth noting that in this work we take the canonical definition of the energy-momentum tensor:
\begin{equation}\begin{split}
T^{\mu\nu} =& \int \frac{\mathrm{d}^4p}{(2\pi)^4} p^\nu \mathcal{V}^\mu \\
	=& \int_p p^\mu p^\nu f_V 
	+ \hbar\,\epsilon^{\mu\lambda\sigma\rho} \int_p \frac{p^\nu p_\lambda n_\sigma}{2\;n \cdot p} \partial_\rho f_A \,,
\end{split}
\end{equation}
which contains quantum correction which is not necessary symmetric.
However, in this appendix we show how to symmetrize the stress tensor without changing any physical observables or the evolution of thermodynamic quantities.
In principle, one can alter the form of the stress tensor by adding divergenceless term
\begin{equation}
T_{\Phi}^{\mu\nu} \equiv T^{\mu\nu} + \frac{1}{2}\partial_{\lambda}(\Phi^{\lambda\mu\nu} + \Phi^{\mu\nu\lambda} + \Phi^{\nu\lambda\mu})\,,
\end{equation}
while the spin density becomes $ S_{\Phi}^{\lambda\mu\nu} \equiv S^{\lambda\mu\nu} - \Phi^{\lambda\mu\nu} $, in order to maintain angular momentum conservation.
Such transformation is referred to as a {\it pseudo-gauge transformation} in Refs.~\cite{Becattini:2012pp,Florkowski:2018fap,Florkowski:2018ahw}, 
In practice, we employ the Schouten identity~\eqref{eq.schouten} and separate the quantum correction of the stress tensor into symmetric and divergence-less anti-symmetric components:
\begin{eqnarray}
&& \hbar \epsilon^{\mu\lambda\sigma\rho} \int_p \frac{p^\nu p_\lambda n_\sigma}{2\;n \cdot p} \partial_\rho f_A \nonumber\\
&=& \frac{\hbar}{2} \int_p \Big(
	\epsilon^{\mu\lambda\sigma\rho} p^\nu + \epsilon^{\nu\lambda\sigma\rho} p^\mu
	 \Big)\frac{ p_\lambda n_\sigma}{2\,n \cdot p}\partial_\rho f_A
\nonumber\\&&\quad
	 + \frac{\hbar}{2} \int_p \Big(
	\epsilon^{\mu\lambda\sigma\rho} p^\nu - \epsilon^{\nu\lambda\sigma\rho} p^\mu
	 \Big)\frac{ p_\lambda n_\sigma}{2\,n \cdot p}\partial_\rho f_A \nonumber\\
&=& \frac{\hbar}{2} \int_p \Big(
	\epsilon^{\mu\lambda\sigma\rho} p^\nu + \epsilon^{\nu\lambda\sigma\rho} p^\mu
	 \Big)\frac{ p_\lambda n_\sigma}{2\,n \cdot p}\partial_\rho f_A
\nonumber\\&&\quad
	 + \frac{\hbar}{2} \int_p \Big(
	\epsilon^{\sigma\rho\mu\nu} p^\lambda + \epsilon^{\rho\mu\nu\lambda} p^\sigma + \epsilon^{\mu\nu\lambda\sigma} p^\rho
	 \Big)\frac{ p_\lambda n_\sigma}{2\,n \cdot p}\partial_\rho f_A \nonumber\\
&=& \frac{\hbar}{2} \int_p \Big(
	\epsilon^{\mu\lambda\sigma\rho} p^\nu + \epsilon^{\nu\lambda\sigma\rho} p^\mu
	 \Big)\frac{ p_\lambda n_\sigma}{2\,n \cdot p}\partial_\rho f_A
\nonumber\\&&\quad
	 + \frac{\hbar}{4} \epsilon^{\mu\nu\lambda\rho} \partial_\lambda \int_p p_\rho f_A  + \mathcal{O}(\hbar^2) \,.
\end{eqnarray}
Especially, the anti-symmetric term vanishes after taking the divergence, $\frac{\hbar}{4} \epsilon^{\mu\nu\lambda\rho} \partial_\mu \partial_\lambda \int_p p_\rho f_A = 0$, and does not contribute to the conservation equation.
This identity also yields the explicit form of the pseudo-gauge transformation:
\begin{equation}
\Phi^{\lambda\mu\nu} \equiv - \frac{\hbar}{6} \epsilon^{\lambda\mu\nu\rho} \int_p p_\rho f_A
\end{equation}
so that
\begin{equation}\begin{split}
T^{\mu\nu}_\mathrm{sym} 
\equiv\,&
	 T^{\mu\nu}_\mathrm{can} 
	+ \frac{1}{2}\partial_{\lambda}(\Phi^{\lambda\mu\nu} + \Phi^{\mu\nu\lambda} + \Phi^{\nu\lambda\mu} ) 
\\=\,&
	\int_p p^\mu p^\nu f_V 
	+ \frac{\hbar}{2} \int_p \Big(
	\epsilon^{\mu\lambda\sigma\rho} p^\nu + \epsilon^{\nu\lambda\sigma\rho} p^\mu
	 \Big)\frac{ p_\lambda n_\sigma}{2\,n \cdot p}\partial_\rho f_A 
\end{split}\end{equation}
is symmetric. Using such a definition, the equilibrium form of stress tensor becomes
\begin{equation}
T_{\mathrm{sym,eq}}^{\mu\nu} 
	= \varepsilon\, u^\mu u^\nu - P\, \Delta^{\mu\nu}
	+ \hbar \,n_A \,(\omega^\mu u^\nu +  \omega^\nu u^\mu ) \,.
\end{equation}
It is worth mentioning that the pseudo-gauge transformation does not bring any ambiguity in our framework, because of the following two reasons. 
First, the additional term is divergenceless by definition, hence it does not alter the evolution of the system. 
Second, although the pseudo-gauge transformation modifies the definition of ``spin density'' $S^{\lambda\mu\nu}$, the spin/chirality dependent distribution function remain the same.
In other words, physical observables in heavy-ion collisions, such as spin polarization vector as shown in Eq.~(\ref{eq.spin_vector}), are independent of the choice of pseudo-gauge.
\section{Thermodynamic Integrals and Orthogonal Polynomials}\label{sec.orthogonal.polynomial}

In this appendix, we discuss some mathematical relations related to the thermodynamics integrals $\int_p (\cdots)f_0$ and $\int_p (\cdots)f_0(1-f_0)$, and construct the orthogonal polynomials used in the main text.

$\bullet$ Integration by Part: In the main text, integration by part is frequently employed to derive/simplify the thermal integrals. Noting that
\begin{equation}\begin{split}
&	\frac{\mathrm{d}}{\mathrm{d}p} f_\mathrm{0} = - \frac{p}{E_p\,T} f_\mathrm{0} (1-f_\mathrm{0}) \,,\\
&	\frac{\mathrm{d}}{\mathrm{d}p} f_\mathrm{0}  (1-f_\mathrm{0}) = - \frac{p}{E_p\,T} f_\mathrm{0} (1-f_\mathrm{0}) (1-2f_\mathrm{0}) \,,
\end{split}\end{equation}
and applying integration by part, one can find
\begin{eqnarray}
&&\begin{split}
&	\int \frac{\mathrm{d}^3\mathbf{p}}{(2\pi)^3\, E_p}  f_\mathrm{0} (1-f_\mathrm{0}) F[E_p, p] 
\\=\,&
	T \int \frac{\mathrm{d}^3\mathbf{p}}{(2\pi)^3\, E_p}  f_\mathrm{0} \frac{E_p}{p^2} \frac{\mathrm{d}}{\mathrm{d}p}( p F[E_p, p] ) \,, \\
\end{split}\\
&&\begin{split}
&	\int \frac{\mathrm{d}^3\mathbf{p}}{(2\pi)^3\, E_p}  f_\mathrm{0} (1-f_\mathrm{0})  (1-2f_\mathrm{0}) F[E_p, p] 
\\=\,&
	T \int \frac{\mathrm{d}^3\mathbf{p}}{(2\pi)^3\, E_p}  f_\mathrm{0} (1-f_\mathrm{0}) \frac{E_p}{p^2} \frac{\mathrm{d}}{\mathrm{d}p}( p F[E_p, p] ) \,.
\end{split}\end{eqnarray}

$\bullet$ Orthogonality in Thermodynamic Integrals:
For an arbitrary function of co-moving energy $F = F(u \cdot p)$, angular dependence yields the orthogonal property:
\begin{eqnarray}\begin{split}
&\int \frac{\mathrm{d}^3\mathbf{p} \; F}{(2\pi)^3\, E_p} p^{\langle \mu_1} \cdots p^{\mu_m\rangle} p_{\langle \nu_1} \cdots p_{\nu_n\rangle} 
\\=\,&
	\frac{m! \delta_{mn}}{(2m+1)!!}\Delta^{\mu_1 \cdots \mu_m}_{\nu_1\cdots\nu_m} \int \frac{\mathrm{d}^3\mathbf{p}\;F}{(2\pi)^3\, E_p} (\Delta^{\alpha\beta}p_\alpha p_\beta)^m \,.
\end{split}
\end{eqnarray}

$\bullet$ Orthogonal Polynomials: we start by defining some thermodynamic integrals as
\begin{eqnarray}
I_{n,q} &\equiv& 
	\int \frac{\mathrm{d}^3\mathbf{p}\, (-\Delta^{\mu\nu} p_\mu p_\nu)^q (u \cdot p)^{n-2q}}
	{(2\pi)^3 E_p\,(2q+1)!! } f_\mathrm{0} \,,\\
J_{n,q} &\equiv& 
	\int \frac{\mathrm{d}^3\mathbf{p}\, (-\Delta^{\mu\nu} p_\mu p_\nu)^q (u \cdot p)^{n-2q}}
	{(2\pi)^3 E_p\,(2q+1)!! } f_\mathrm{0} (1-f_\mathrm{0}) \,,\qquad\\
G_{n,m}^{(q)} &\equiv& J_{n,q} J_{m,q} - J_{n-1,q} J_{m+1,q} \,,\\
G_{n,m} &\equiv& G_{n,m}^{(0)} = J_{n,0} J_{m,0} - J_{n-1,0} J_{m+1,0} \,,\\
D_{n,q} &\equiv& J_{n+1,q} J_{n-1,q} - J_{n,q}^2 \,.
\end{eqnarray}
Then we construct the polynomials $P^{(\ell)}_m$ as functions of the co-moving energy $E_p \equiv (u\cdot p)$.
They are defined to satisfy the orthonormal relation:
\begin{eqnarray}
\delta_{mn} = \int \frac{\mathrm{d}^3\mathbf{p}}{(2\pi)^3\, E_p} \omega^{(\ell)} P^{(\ell)}_m  P^{(\ell)}_n \,,
\end{eqnarray}
where the weight function
\begin{eqnarray}
\omega^{(\ell)} = \frac{(-1)^{(\ell)} ( \Delta^{\mu\nu} p_\mu p_\nu )^\ell}{(2\ell+1)!! J_{2\ell,\ell}} f_\mathrm{0}(p) (1-f_\mathrm{0}(p))\,,
\end{eqnarray}
satisfies the normalization relation
\begin{eqnarray}
1 	= \int \frac{\mathrm{d}^3\mathbf{p}}{(2\pi)^3\, E_p} \omega^{(\ell)} 
	\,.
\end{eqnarray}
For each $\ell$, we explicitly write down the $0^\mathrm{th}$-,  $1^\mathrm{st}$-, and $2^\mathrm{nd}$-order polynomials as
\begin{eqnarray}
&&P_0^{(\ell)} = 1 \,,\\
&&P_1^{(\ell)} = \frac{J_{2\ell+1,\ell}}{\sqrt{D_{2\ell+1,\ell}}} - \frac{J_{2\ell,\ell}}{\sqrt{D_{2\ell+1,\ell}}}(u\cdot p) \,,\\
&&P_2^{(\ell)} = \frac{D_{2\ell+2,\ell} - G^{(\ell)}_{2\ell+3,2\ell}(u\cdot p) + D_{2\ell+1,\ell}(u\cdot p)^2}{\sqrt{N_\ell}} \,.\nonumber\\
\end{eqnarray}
where the normalization factor is
\begin{equation}\begin{split}
N_\ell \equiv& \frac{D_{2\ell+1,\ell}}{J_{2\ell,\ell}} \Big(
 	J_{2\ell+2,\ell}D_{2\ell+2,\ell} \\
&\qquad	- J_{2\ell+3,\ell} G^{(\ell)}_{2\ell+3,2\ell} 
	+ J_{2\ell+4,\ell}D_{2\ell+1,\ell} \Big) \,.
\end{split}
\end{equation}

We further define
\begin{equation}
\mathcal{F}^{[X],\pm}_{r,q} \equiv
	\frac{(-1)^q q!}{(2q+1)!!}\int_p f_{\mathrm{0},\pm} (1-f_{\mathrm{0},\pm})
	 \frac{(-\Delta_{\alpha \beta} p^\alpha p^\beta)^q}{(u\cdot p)^r} \lambda_{X}^\pm \,,
\end{equation}
with $X$ being $\Pi$, $\nu$, $\pi$, or $\Omega$.
In particular, matching relations ensures that
\begin{equation}\begin{split}
&	\mathcal{F}^{[\Pi],\pm}_{0,0} = -\frac{3}{2m^2},\quad
	\mathcal{F}^{[\Pi],\pm}_{-1,0} = 0,\quad
	\mathcal{F}^{[\Pi],\pm}_{-2,0} = 0,
\\&	\mathcal{F}^{[\pi],\pm}_{0,2} = 1/2,\quad
	\mathcal{F}^{[\nu],\pm}_{0,1} = 1,\quad
	\mathcal{F}^{[\nu],\pm}_{-1,1} = 0,
\\&	\mathcal{F}^{[\Omega],\pm}_{1,1} = 0,\qquad
	\mathcal{F}^{[\Omega],\pm}_{0,1} = -1.
\end{split}\end{equation}
Similarly, we have
\begin{equation}\begin{split}
&	I_{1,0}^\pm = J_{2,1}^{\pm}/T = n_\pm \,,\qquad
	I_{2,0}^\pm = \epsilon_\pm \,,\\
&	J_{3,1}^{\pm} = T(\epsilon_\pm + P_\pm) \,,\qquad
	J_{1,0}^\pm = \frac{\partial n^\pm}{\partial\alpha^\pm}\,.
\end{split}\end{equation}
From the definition and after integration by parts, one can find
\begin{eqnarray}
&&	J_{n,q} = \frac{\partial I_{n,q}}{\partial \alpha}\Big|_\beta\,,\\
&&	J_{n,q} = -\frac{\partial I_{n-1,q}}{\partial \beta}\Big|_\alpha\,,\\
&&	J_{n,q} = (n+1)T\,I_{n-1,q}\,.
\end{eqnarray}

$\bullet$ Simplification of Thermodynamic Integrals: Employing the on-shell condition $( - \Delta^{\mu\nu} p_\mu p_\nu) = (u \cdot p)^2 - m^2$, one can find
\begin{eqnarray}
I_{n,q} &=& \frac{q!}{(2q+1)!!}\sum_{k=0}^{q} \frac{(-1)^k m^{2k}}{k! (q-k)!} I_{n-2k,0} \,,\\
J_{n,q} &=& \frac{q!}{(2q+1)!!}\sum_{k=0}^{q} \frac{(-1)^k m^{2k}}{k! (q-k)!} J_{n-2k,0} \,,\\
\mathcal{F}^{[X],\pm}_{r,q} &=&  \frac{(-1)^q (q!)^2}{(2q+1)!!}\sum_{k=0}^{q} \frac{(-1)^k m^{2k}}{k! (q-k)!} \mathcal{F}^{[X],\pm}_{r+2k-2q,0}\,.\qquad
\end{eqnarray}
These expressions can be further simplified when taking the massless limit $m=0$, 
\begin{eqnarray}
I_{n,q} &=& \frac{1}{(2q+1)!!}  I_{n,0} \,,\\
J_{n,q} &=& \frac{1}{(2q+1)!!} J_{n,0} \,,\\
D_{n,q} &=& \Big[\frac{1}{(2q+1)!!}\Big]^2 D_{n,0} \,,\\
G_{n,m}^{(q)} &=& \Big[\frac{1}{(2q+1)!!}\Big]^2 G_{n,m}\,,\\
\mathcal{F}^{[X],\pm}_{r,q} &=&  \frac{(-1)^q q!}{(2q+1)!!} \mathcal{F}^{[X],\pm}_{r-2q,0}\,.
\end{eqnarray}

\section{Coefficients in Dissipative Quantities}\label{appendix.moment_coefficients}
In this appendix, we show the full details of computing the coefficients $\lambda_X$ obtained from matching dissipative quantities with non-equilibrium distribution functions.
In the moment expansion formalism, we expand the distribution functions near their equilibrium forms:
\begin{equation}\begin{split}
f^\pm 
\equiv\,& f_\mathrm{eq}^\pm + f_\mathrm{eq}^\pm (1-f_\mathrm{eq}^\pm)
\bigg[ 
\\&\quad
+	\lambda_{\Pi}^\pm \Pi
+	\lambda_{\nu}^\pm \nu_{\pm}^\mu p_\mu 
+	\lambda_{\pi}^\pm \pi^{\mu\nu} p_\mu p_\nu 
\bigg] \,,
\end{split}\end{equation}
where the non-equilibrium corrections can be expressed as
\begin{eqnarray}
\lambda_{\Pi}^\pm \Pi &=&
	 c_{\pm,0} P^{(0)}_0 + c_{\pm,1} P^{(0)}_1 + c_{\pm,2} P^{(0)}_2  \,,\label{eq.lambda_Pi}\\
\lambda_{\nu}^\pm \nu_\pm^{\alpha} &=&
	 c_{\pm,0}^\alpha P^{(1)}_0 + c_{\pm,1}^\alpha P^{(1)}_1 \,,\\
\lambda_{\pi}^\pm \pi^{\alpha\beta} &=&
	 c_{\pm,0}^{\alpha\beta} P^{(2)}_0 \,,\label{eq.lambda_pi}
\end{eqnarray}
In above equations, $P^{(\ell)}_n$ are orthogonal polynomials of co-moving energy $(u \cdot p)$, and their explicit form can be found in Sec.~\ref{sec.orthogonal.polynomial}.
Additionally, $(c_{\pm,0}, c_{\pm,1}, c_{\pm,2}, c_{\pm,0}^\alpha, c_{\pm,1}^\alpha, c_{\pm,0}^{\alpha\beta})$ are coefficients that depend on temperature $T$, chemical potential $\mu^{\pm}$, fluid velocity $u^\mu$, but not on momentum $p$. 
In addition, the coefficients are orthogonal to velocity:
\begin{equation}
c_{\pm}^{\mu} \equiv \Delta^{\mu}_{\alpha} c_{\pm}^{\alpha} \,, \quad
c_{\pm}^{\mu\nu} \equiv \Delta^{\mu\nu}_{\alpha\beta} c_{\pm}^{\alpha\beta}  \,.
\end{equation}
It might be worth mentioning that although it has been shown in the main text that $\Pi$ as well as the scalar correction $\lambda_\Pi \Pi$ vanish for massless system, we formally keep these terms in this appendix, for the convenience of future extensions.

To determine the coefficients, we first denote $\delta f_\pm \equiv f^\pm - f_\mathrm{eq}^\pm$, and compute the integrals:
\begin{eqnarray}
&&\int_p \delta f_\pm 
	= J_{0,0}^{\pm} c_{\pm,0} \,,\\~\nonumber\\
&&\int_p (u\cdot p) \delta f_\pm 
	= J_{1,0}^{\pm} c_{\pm,0} - \sqrt{D_{1,0}^{\pm}} c_{\pm,1}\,,\\~\nonumber\\
&&\begin{split}
&	\int_p (u\cdot p)^2 \delta f_\pm 
	= J_{2,0}^{\pm} c_{\pm,0}  -  \frac{G_{3,0}^{\pm}}{\sqrt{D_{1,0}^{\pm}}}c_{\pm,1} \\
&\qquad
	+ \frac{\sqrt{J_{2,0}^{\pm} D_{2,0}^{\pm} - J_{3,0}^{\pm} G_{3,0}^{\pm} + J_{4,0}^{\pm} D_{1,0}^{\pm}}}
		{\sqrt{D_{1,0}^{\pm} / J_{0,0}^{\pm}}}c_{\pm,2} \,,\qquad
\end{split}
\\~\nonumber\\
&&\begin{split}
&\int_p \Delta^{\mu\alpha}p_\alpha \delta f_\pm 
	= - J_{2,1}^{\pm} c_{\pm,0}^\mu
	\,,
\end{split}\\~\nonumber\\
&&\begin{split}
&\int_p (u\cdot p)\Delta^{\mu\alpha}p_\alpha \delta f_\pm 
	= - J_{3,1}^{\pm} c_{\pm,0}^\mu
	+ \sqrt{D_{3,1}^{\pm}} c_{\pm,1}^\mu
	\,,\qquad
\end{split}\\~\nonumber\\
&&\int_p \Delta^{\mu\nu}_{\alpha\beta}p^\alpha p^\beta \delta f_\pm 
	= 2J_{4,2}^{\pm} c_{\pm,0}^{\mu\nu} \,.
\end{eqnarray}
Keeping up to $\hbar^0$-order, we find
\begin{eqnarray}
&&\begin{split}
&\int_p \frac{p^\mu}{u \cdot p} f_{\pm}(p) =
	- \Big( J_{1,1}^{\pm} c_{\pm,0}^\mu + \frac{D_{2,1}^{\pm}}{\sqrt{D_{3,1}^{\pm}}}c_{\pm,1}^\mu \Big)
\\&\qquad
	+\Big( I_{0,0}^{\pm} + J_{0,0}^{\pm} c_{\pm,0}\Big) u^\mu \,,
\end{split}\\~\nonumber\\
&&\begin{split}
&\int_p \frac{p^\mu p^\nu}{(u \cdot p)^2} f_{\pm}(p) 
	= I_{0,0}^{\pm} u^\mu u^\nu - I_{0,1}^{\pm} \Delta^{\mu\nu}
\\&\qquad
	 - J_{1,1}^{\pm} (u^\mu c_{\pm,0}^\nu + u^\nu c_{\pm,0}^\mu)
	+ 2J_{2,2}^{\pm} c_{\pm,0}^{\mu\nu} \,,
\end{split}\\~\nonumber\\
&&\begin{split}
& \int_p \frac{p^\mu p^\nu}{u \cdot p} f_{\pm}(p) 
	= I_{1,0}^{\pm} u^\mu u^\nu-I_{1,1}^{\pm} \Delta^{\mu\nu}  
\\&\qquad
	 - J_{2,1}^{\pm} (u^\mu c_{\pm,0}^\nu   +u^\nu c_{\pm,0}^\mu) 
	+ 2J_{3,2}^{\pm} c_{\pm,0}^{\mu\nu} \,,
\end{split}\\~\nonumber\\
&&\begin{split}
& \int_p \frac{p^{\langle \mu \rangle} p^{\langle \nu \rangle} p^\lambda}{(u \cdot p)^2} f_{\pm}(p) 
	= -I_{1,1}^{\pm} \Delta^{\mu\nu} u^\lambda 
	+ 2 J_{3,2}^{\pm} c_{\pm,0}^{\mu\nu} u^{\lambda} 
\\&\qquad
	 + \bigg(\Delta^{\mu \nu} \Delta^{\lambda}_{\alpha} +\Delta^{\mu \lambda} \Delta^{\nu}_{\alpha} +\Delta^{\lambda \nu} \Delta^{\mu}_{\alpha}  \bigg)
\\&\qquad\qquad\times
	 \bigg( J_{2,2}^\pm c_{\pm,0}^{\alpha} + \frac{J_{3,1}^\pm J_{2,2}^\pm -J_{2,1}^\pm J_{3,2}^\pm }{\sqrt{D_{3,1}^\pm}}c_{\pm,1}^{\alpha}
	\bigg)\,.
\end{split}
\end{eqnarray}
Then, the matching relations of Eqs.~(\ref{eq.definition_epsilon} - \ref{eq.definition_nu}) require
\begin{equation}\begin{split}
&
c_{\pm,0} = -\frac{3\Pi}{2m^2 J_{0,0}^\pm} \,,\qquad
c_{\pm,1} = \frac{J_{1,0}^\pm}{\sqrt{D_{1,0}^\pm}} c_{\pm,0} \,,\\
&
c_{\pm,2} = \frac{D_{2,0}^\pm \sqrt{J_{0,0}^\pm/D_{1,0}^\pm}}{\sqrt{J_{2,0}^\pm D_{2,0}^\pm - J_{3,0}^\pm G_{3,0}^\pm + J_{4,0}^\pm D_{1,0}^\pm}} c_{\pm,0} \,,\\
&
c_{\pm,0}^\mu = -\frac{\nu_{\pm}^\mu}{J_{2,1}^\pm} \,,\qquad
c_{\pm,1}^\mu = \frac{J_{3,1}^\pm}{\sqrt{D_{3,1}^\pm}} c_{\pm,0}^\mu\,, \\
& c_{\pm,0}^{\mu\nu} = \frac{\pi^{\mu\nu}}{4J_{4,2}^\pm } \,.
\end{split}\end{equation}
Finally, substituting the coefficients in Eqs.(\ref{eq.lambda_Pi} - \ref{eq.lambda_pi}), one eventually obtain:
\begin{eqnarray}
\lambda_{\Pi}^\pm &\equiv&
	-\frac{3}{2m^2 J_{0,0}^\pm} \Bigg( P^{(0),\pm}_0 +  \frac{J_{1,0}^\pm}{\sqrt{D_{1,0}^\pm}} P^{(0),\pm}_1 
\nonumber\\&&\quad
	+ \frac{D_{2,0}^\pm \sqrt{J_{0,0}^\pm/D_{1,0}^\pm}  \;\; P^{(0),\pm}_2}
	{\sqrt{J_{2,0}^\pm D_{2,0}^\pm - J_{3,0}^\pm G_{3,0}^\pm + J_{4,0}^\pm D_{1,0}^\pm}} \Bigg) \,,\quad 
\\
\lambda_{\nu}^\pm &\equiv&
	-\frac{1}{J_{2,1}^\pm} \Bigg(P^{(1),\pm}_0  + \frac{J_{3,1}^\pm P^{(1),\pm}_1}{\sqrt{D_{3,1}^\pm}}  \Bigg) 
\nonumber\\&=&
	 \frac{J_{3,1}^\pm (u \cdot p) - J_{4,1}^\pm}{D_{3,1}^\pm} \,,\\
 \lambda_{\pi}^\pm &\equiv&
	\frac{P^{(2),\pm}_0}{4J_{4,2}^\pm} 
=	\frac{1}{4J_{4,2}^\pm} \,.
\end{eqnarray}

With these, we have
\begin{eqnarray}
&& \mathcal{F}^{[\pi],\pm}_{r,q} =
	(-1)^q q! \frac{J_{2q-r,q}^{\pm}}{4J_{4,2}^{\pm}} \,,\\
&&\mathcal{F}^{[\nu],\pm}_{r,q} =
	(-1)^q q! \frac{J_{3,1}^\pm J_{2q-r+1,q}^{\pm} - J_{4,1}^\pm J_{2q-r,q}^{\pm}}{D_{3,1}^{\pm}} \,.\qquad
\end{eqnarray}

\section{Other Mathematical Relations}\label{sec.mathematical_relations}

In this appendix, we list some of the mathematical relations employed in the derivation.
 
$\bullet$ Schouten identity --- in this paper, we frequently employ the following identity:
\begin{equation}
0 = p^\mu \epsilon^{\nu\rho\sigma\lambda} + p^\nu \epsilon^{\rho\sigma\lambda\mu}  + p^\rho \epsilon^{\sigma\lambda\mu\nu}  + p^\sigma \epsilon^{\lambda\mu\nu\rho}  + p^\lambda \epsilon^{\mu\nu\rho\sigma} \,.
\label{eq.schouten}
\end{equation}

$\bullet$ Projector:
\begin{eqnarray}
\Delta^{\alpha\beta\gamma}_{\mu\nu\lambda} &=&
	 \frac{1}{6}\Big(
	\Delta^{\alpha}_{\mu}\Delta^{\beta}_{\nu}\Delta^{\gamma}_{\lambda} 
	+ \Delta^{\alpha}_{\nu}\Delta^{\beta}_{\lambda}\Delta^{\gamma}_{\mu} 
	+ \Delta^{\alpha}_{\lambda}\Delta^{\beta}_{\mu}\Delta^{\gamma}_{\nu} 
\nonumber\\&&\qquad
	+ \Delta^{\alpha}_{\mu}\Delta^{\beta}_{\lambda}\Delta^{\gamma}_{\nu} 
	+ \Delta^{\alpha}_{\nu}\Delta^{\beta}_{\mu}\Delta^{\gamma}_{\lambda} 
	+ \Delta^{\alpha}_{\lambda}\Delta^{\beta}_{\nu}\Delta^{\gamma}_{\mu} 
	 \Big)
\nonumber\\&&
	-\frac{1}{15}\Big(
	\Delta^{\alpha\beta}\Delta_{\mu\nu}\Delta^{\gamma}_{\lambda} 
	+\Delta^{\alpha\beta}\Delta_{\nu\lambda}\Delta^{\gamma}_{\mu} 
	+\Delta^{\alpha\beta}\Delta_{\lambda\mu}\Delta^{\gamma}_{\nu} 
\nonumber\\&&\qquad
	+\Delta^{\beta\gamma}\Delta_{\mu\nu}\Delta^{\alpha}_{\lambda} 
	+\Delta^{\beta\gamma}\Delta_{\nu\lambda}\Delta^{\alpha}_{\mu} 
	+\Delta^{\beta\gamma}\Delta_{\lambda\mu}\Delta^{\alpha}_{\nu} 
\nonumber\\&&\qquad
	+\Delta^{\gamma\alpha}\Delta_{\mu\nu}\Delta^{\beta}_{\lambda} 
	+\Delta^{\gamma\alpha}\Delta_{\nu\lambda}\Delta^{\beta}_{\mu} 
	+\Delta^{\gamma\alpha}\Delta_{\lambda\mu}\Delta^{\beta}_{\nu} 
	 \Big)\,.\nonumber\\
\end{eqnarray}

$\bullet$ Simplifying quantum correction term in CKE:
\begin{eqnarray}
&&	\hbar \delta(p^2) \Big(\partial_\mu \frac{\epsilon^{\mu\nu\rho\sigma}p_\rho u_\sigma}{2 p \cdot u}\Big)\partial_\nu f \nonumber\\
&=&	\hbar \delta(p^2) \Big(
	\frac{\epsilon^{\mu\nu\rho\sigma}p_\rho \partial_\mu u_\sigma}{2 p \cdot u} 
	- \frac{\epsilon^{\mu\nu\rho\sigma}p^\lambda p_\rho u_\sigma \partial_\mu u_\lambda}{2 (p \cdot u)^2} 
	\Big)\partial_\nu f \nonumber\\
&=&	\hbar \delta(p^2) \Big(
	\frac{\epsilon^{\mu\nu\rho\sigma} p_\rho \partial_\mu u_\sigma}{2 p \cdot u} 
	- \frac{\epsilon^{\mu\nu\rho\sigma} p^\lambda p_\rho u_\sigma (\partial_\mu u_\lambda + \partial_\lambda u_\mu)}{4 (p \cdot u)^2} 
\nonumber\\&&\qquad\qquad
	- \frac{\epsilon^{\mu\nu\rho\sigma} p^\lambda p_\rho u_\sigma (\partial_\mu u_\lambda - \partial_\lambda u_\mu)}{4 (p \cdot u)^2} 
	\Big)\partial_\nu f \nonumber\\
&=&	\hbar \delta(p^2) \Big(
	\frac{\epsilon^{\mu\nu\rho\sigma} p_\rho\partial_{[\mu}u_{\sigma]}}{2 p \cdot u} 
	- \frac{\epsilon^{\mu\nu\rho\sigma} p^\lambda p_\rho u_\sigma \partial_{[\mu}u_{\lambda]}}{2 (p \cdot u)^2} 
	\Big)\partial_\nu f \nonumber\\
&=&	\hbar \delta(p^2) \Big(
	\frac{\epsilon^{\mu\nu\rho\sigma} p_\rho \partial_{[\mu}u_{\sigma]}}{2 p \cdot u} 
	+ (-\epsilon^{\mu\nu\rho\sigma} p^\lambda - \epsilon^{\lambda\nu\rho\sigma} p^\mu
\nonumber\\&&\qquad\qquad
	+\epsilon^{\rho\sigma\lambda\mu} p^\nu + \epsilon^{\sigma\lambda\mu\nu} p^\rho  + \epsilon^{\lambda\mu\nu\rho} p^\sigma)
	 \frac{p_\rho u_\sigma \partial_{[\mu}u_{\lambda]}}{4 (p \cdot u)^2} 
	\Big)\partial_\nu f \nonumber\\
&=&\label{eq.quantum_correction_in_CKE}
	\hbar \delta(p^2) \Big(
	\frac{\epsilon^{\mu\nu\rho\sigma} p_\nu(\partial_\rho u_\sigma) }{4 p \cdot u} 
	\Big)\partial_\mu f + \mathcal{O}(\hbar^2)\,.
\end{eqnarray}

\clearpage
\begin{widetext}

\section{Equation of motion for Dissipative Quantities}\label{app.eom_dissipative}

In this appendix, we derive the equations of motion for dissipative terms, ruled by:
\begin{eqnarray}
\Delta^{\mu\nu}_{\rho\sigma} \hat{\mathrm{d}}\pi^{\rho\sigma} &\equiv& 
	\int_p \Delta^{\mu\nu}_{\alpha\beta} p^\alpha p^\beta \Big( \hat{\mathrm{d}} \delta f_{+} + \hat{\mathrm{d}} \delta f_{-}\Big) \,, \\
\Delta^{\mu\nu} \hat{\mathrm{d}} \nu_{\pm,\nu} &\equiv& \int_p \Delta^\mu_\alpha p^\alpha \hat{\mathrm{d}}\delta  f_\pm \,.
\end{eqnarray}
where $\delta f_\pm \equiv f^\pm - f_\mathrm{eq}^\pm$, and
\begin{equation}\begin{split}
& \hat{\mathrm{d}} \delta f_\pm 
	- \Big( \frac{1}{u\cdot p} \pm \hbar \frac{\omega\cdot p}{2\, (u \cdot p)^3} \Big) \mathcal{C}_\pm[f_+, f_-] 
=
	- \hat{\mathrm{d}} f_{\mathrm{eq},\pm} 
 	- \frac{p^\mu \nabla_\mu f_\pm}{u\cdot p}
	\mp \frac{\hbar  \epsilon^{\mu \nu \lambda \sigma} p_\nu p^\rho u_\lambda (\partial_\rho u_\sigma - \partial_\sigma u_\rho)}{4(u \cdot p)^3} \nabla_\mu f_\pm\,.
\end{split}
\end{equation}
Although it has been proven that the bulk viscous pressure $\Pi$ vanishes for massless system, we keep it for later convenience
\begin{eqnarray}
&&\begin{split}
f^\pm =\,& f_0^\pm + f_0^\pm (1-f_0^\pm)
	\bigg[ 
		\mp	\frac{\hbar}{2T}\frac{\omega \cdot p}{u\cdot p}		+	\lambda_{\Pi}^\pm \Pi
		+	\lambda_{\nu}^\pm \nu_{\pm}^\mu p_\mu 
		+	\lambda_{\pi}^\pm \pi^{\mu\nu} p_\mu p_\nu 
	\bigg] \,,
\end{split}\nonumber\\~\\
&&\begin{split}
&\lambda_{\pi}^\pm = \frac{1}{4J_{4,2}^\pm} \,,\qquad
\lambda_{\nu}^\pm =  \frac{J_{3,1}^\pm (u \cdot p) - J_{4,1}^\pm}{D_{3,1}^\pm} \,.
\end{split}
\end{eqnarray}

From conservation equations one can find:
\begin{eqnarray}
-\hat{\mathrm{d}} n_{\pm} 
	&=& n_{\pm}\theta + \partial_\mu \nu_{\pm}^\mu
	\pm \hbar\, \partial_\mu ( I_{0,0}^\pm \omega^\mu) \,,
\\
	-\hat{\mathrm{d}}\epsilon
&=&
	 (\epsilon+P)\theta 
	 - \pi^{\alpha\beta}\sigma_{\alpha\beta}
	+ \frac{\hbar}{2} n_A u_\nu \hat{\mathrm{d}} \omega^\nu 
	+ \frac{3\hbar}{2} \partial_\mu (  n_A \omega^\mu) \,,
\\
-\hat{\mathrm{d}}u^\nu &=&
	\frac{1}{\epsilon + P}\Big(
	 - \nabla^{\nu}P 
	 + \Delta^\nu_{\alpha} \partial_{\beta}\pi^{\alpha\beta}
	+ \frac{\hbar}{2} n_A \Delta^\nu_{\alpha} \hat{\mathrm{d}} \omega^\alpha
	+ \frac{3\hbar}{2} n_A \omega^\mu \nabla_\mu u^\nu
	\Big) \,.
\end{eqnarray}

Then, we obtain the equation of motion for the shear viscous tensor:
\begin{eqnarray}
&& 
	\Delta^{\alpha\beta}_{\rho\sigma} \hat{\mathrm{d}}\pi^{\rho\sigma} 
	- (\mathcal{A}_{+,0}^{(2)} + \mathcal{A}_{-,0}^{(2)}) \pi^{\alpha\beta}	
	- \frac{\hbar}{2} (\mathcal{X}_{2,-2}^{+,+} - \mathcal{X}_{2,-2}^{-,+})
		 \Delta^{\alpha\beta}_{\rho\sigma} \omega^\rho \nu_+^\sigma
	+ \frac{\hbar}{2} (\mathcal{X}_{2,-2}^{-,-} -  \mathcal{X}_{2,-2}^{+,-}) 
		\Delta^{\alpha\beta}_{\rho\sigma} \omega^\rho \nu_-^\sigma 
\nonumber\\&=& 
	- \int_p p^{\langle\alpha} p^{\beta\rangle} \hat{\mathrm{d}} f_{\mathrm{eq},+} 
	- \int_p \frac{p^{\langle\alpha} p^{\beta\rangle}  p^\mu}{u\cdot p}  \nabla_\mu f_+
	- \frac{\hbar}{4} \epsilon^{\mu \nu \lambda \sigma} u_\lambda (\partial_\rho u_\sigma - \partial_\sigma u_\rho) 
		\int_p \frac{p^{\langle\alpha} p^{\beta\rangle}  p_\nu p^\rho }{(u \cdot p)^3}\nabla_\mu f_+
\nonumber\\&&
	- \int_p p^{\langle\alpha} p^{\beta\rangle} \hat{\mathrm{d}} f_{\mathrm{eq},-} 
	- \int_p \frac{p^{\langle\alpha} p^{\beta\rangle}  p^\mu}{u\cdot p}  \nabla_\mu f_-
	+ \frac{\hbar}{4} \epsilon^{\mu \nu \lambda \sigma} u_\lambda (\partial_\rho u_\sigma - \partial_\sigma u_\rho) 
		\int_p \frac{p^{\langle\alpha} p^{\beta\rangle}  p_\nu p^\rho }{(u \cdot p)^3}\nabla_\mu f_- 
\nonumber\\&=& 
	\frac{8}{5} P \sigma^{\alpha\beta}
	- 3\theta\,\pi^{\alpha\beta} 
	+ \frac{8}{7} \Delta^{\alpha\beta} \sigma^{\mu\nu} \pi_{\mu\nu} 
	-\frac{12}{7} \sigma^{\alpha}_{\;\;\mu} \pi^{\beta\mu}
	-\frac{12}{7} \sigma^{\beta}_{\;\;\mu} \pi^{\alpha\mu}
	-\pi^{\alpha}_{\;\;\mu}\epsilon^{\beta\mu\nu\rho} u_\nu \omega_\rho 
	- \pi^{\beta}_{\;\;\mu}\epsilon^{\alpha\mu\nu\rho} u_\nu \omega_\rho
\nonumber\\&&
	+ \frac{2\hbar}{15} \Delta^{\alpha\beta}_{\mu\nu} \omega^\mu \nabla^\nu n_A 
	+ \frac{\hbar}{5}  n_A \,\Delta^{\alpha\beta}_{\mu\nu} \nabla^{\mu} \omega^\nu 
	- \frac{9\hbar}{10}n_A\, \Delta^{\alpha\beta}_{\mu\nu} \omega^\mu \hat{\mathrm{d}} u^\nu
\nonumber\\&&
	+ \frac{\hbar\,n_A}{20} \Big[\sigma_{\;\;\mu}^{\beta} 
		\epsilon^{\mu \alpha \lambda \sigma} u_\lambda (\hat{\mathrm{d}} u_\sigma) 
	+\sigma_{\;\;\mu}^{\alpha} \epsilon^{\mu \beta \lambda \sigma} u_\lambda (\hat{\mathrm{d}} u_\sigma) \Big]
\nonumber\\&=& 
	\frac{8}{5} P \sigma^{\alpha\beta}
	- 3\theta\,\pi^{\alpha\beta} 
	+ \frac{8}{7} \Delta^{\alpha\beta} \sigma^{\mu\nu} \pi_{\mu\nu} 
	-\frac{12}{7} \sigma^{\alpha}_{\;\;\mu} \pi^{\beta\mu}
	-\frac{12}{7} \sigma^{\beta}_{\;\;\mu} \pi^{\alpha\mu}
	-\pi^{\alpha}_{\;\;\mu}\epsilon^{\beta\mu\nu\rho} u_\nu \omega_\rho 
	-\pi^{\beta}_{\;\;\mu}\epsilon^{\alpha\mu\nu\rho} u_\nu \omega_\rho
\nonumber\\&&
	+ \frac{2\hbar}{15} \Delta^{\alpha\beta}_{\mu\nu} \omega^\mu \nabla^\nu n_A 
	+ \frac{\hbar}{5}  n_A \,\Delta^{\alpha\beta}_{\mu\nu} \nabla^{\mu} \omega^\nu 
	- \frac{9\hbar}{10}\frac{n_A}{\varepsilon+P} \Delta^{\alpha\beta}_{\mu\nu} \omega^\mu \nabla^\nu P
\nonumber\\&&
	+ \frac{\hbar}{20}\frac{n_A}{\varepsilon+P} \Big( \sigma_{\;\;\mu}^{\beta} 
		\epsilon^{\mu \alpha \lambda \sigma} u_\lambda \nabla_\sigma P
	+\sigma_{\;\;\mu}^{\alpha} \epsilon^{\mu \beta \lambda \sigma} u_\lambda \nabla_\sigma P \Big)\,,
\end{eqnarray}
for dissipative currents:
\begin{eqnarray}
&&
	\Delta^{\alpha\beta} \hat{\mathrm{d}}\nu_{\beta} ^{\pm}
	- \mathcal{A}_{\pm,0}^{(1)} \nu^\alpha_\pm
	- \mathcal{B}_{\pm,0}^{(1)} \nu^\alpha_\mp
	\pm \frac{\hbar}{2T} \mathcal{W}_{\pm,0}^{(1)} \omega^\alpha
	\pm \hbar \mathcal{U}_{\pm,0}^{(1)} \Omega_+^\alpha
	\pm \hbar \mathcal{V}_{\pm,0}^{(1)} \Omega_-^\alpha
	+ \frac{\hbar}{2} \Big(\mathcal{A}_{+,-2}^{(2)} - \mathcal{A}_{-,-2}^{(2)}\Big) \pi^{\alpha\beta}\omega_\alpha
\nonumber\\	&=& 
	- \int_p p^{\langle\alpha\rangle} \hat{\mathrm{d}} f_{\mathrm{eq},\pm} 
	- \int_p \frac{p^{\langle\alpha\rangle} p^\mu}{u\cdot p}  \nabla_\mu f_\pm
	\mp \frac{\hbar}{4} \epsilon^{\mu \nu \lambda \sigma} u_\lambda (\partial_\rho u_\sigma - \partial_\sigma u_\rho) 
	\int_p \frac{p^{\langle\alpha\rangle} p_\nu p^\rho }{(u \cdot p)^3}\nabla_\mu f_\pm
\nonumber\\	&=&
 	\Big[- n_\pm \hat{\mathrm{d}} u^\alpha
	\mp  \frac{\hbar}{3} \Delta^\alpha_\beta \hat{\mathrm{d}} (I_{0,0}^{\pm}\omega^\beta)\Big]
	+\Big[\frac{1}{3}\nabla^\alpha n_\pm
	- \frac{J_{3,0}^\pm}{2J_{4,0}^\pm} \Delta^{\alpha}_{\rho} \nabla_\mu \pi^{\mu\rho}
	-\pi^{\alpha\mu} \nabla_\mu  \frac{J_{3,0}^\pm}{2J_{4,0}^\pm}
	- \theta \nu_\pm^\alpha 
	- \frac{3}{5}\sigma^{\alpha\mu} \nu^\pm_\mu
\nonumber\\&&
	- \epsilon^{\alpha\mu\nu\gamma} u^\mu \nu_\pm^\nu \omega^\gamma
	\mp\frac{\hbar}{3} I_{0,0}^\pm \Big(\theta \omega^\alpha + \frac{3}{5}\sigma^{\alpha\mu} \omega_\mu \Big)\Big]
	\pm\Big[\frac{\hbar}{12} \epsilon^{\mu \alpha \lambda \sigma} u_\lambda 
		\hat{\mathrm{d}} u_\sigma \nabla_\mu I_{0,0}^\pm
	- \frac{\hbar}{15} I_{0,0}^\pm \Big(\sigma^{\alpha\mu}\omega_\mu - \frac{2\theta}{3}\omega^\alpha \Big)
	\Big]
\nonumber\\	&=&
	\frac{D_{2,1}^\pm}{J_{3,1}^\pm}\nabla^\alpha \frac{\mu_\pm}{T}
	+\ \frac{D_{3,0}^\pm}{2J_{3,0}^\pm J_{4,0}^\pm} \Delta^{\alpha}_{\rho} \nabla_\mu \pi^{\mu\rho}
	-\pi^{\alpha\mu} \nabla_\mu  \frac{J_{3,0}^\pm}{2J_{4,0}^\pm}
	-  \theta \nu_\pm^\alpha - \frac{3}{5}\sigma^{\alpha\mu} \nu^\pm_\mu  
	- \epsilon^{\alpha\mu\nu\gamma} u^\mu \nu_\pm^\nu \omega^\gamma 
	\mp  \frac{\hbar}{3} \omega^\alpha \hat{\mathrm{d}} I_{0,0}^{\pm}
\nonumber\\&&
	\mp \frac{\hbar}{2T} \frac{D_{2,1}^\pm}{J_{3,1}^\pm} \Delta^\alpha_{\beta} \hat{\mathrm{d}} \omega^\beta
	\pm \frac{3\hbar}{2} \frac{n_\pm^2}{\varepsilon_\pm+P_\pm} \Big( \frac{1}{3}\theta \omega^\alpha + \sigma^{\alpha\mu} \omega_\mu \Big)
	\mp\frac{\hbar}{3} I_{0,0}^\pm \Big( \frac{13}{15}\theta \omega^\alpha + \frac{4}{5}\sigma^{\alpha\mu} \omega_\mu \Big)
	\pm \frac{\hbar}{12} \epsilon^{\mu \alpha \lambda \sigma} u_\lambda \hat{\mathrm{d}} u_\sigma (\nabla_\mu I_{0,0}^\pm)\,,
\nonumber\\
\end{eqnarray}

The following relations are useful in the calculations above

\begin{eqnarray}
&&	 \int_p p^\alpha f_\pm
=	n_\pm u^\alpha + \nu_\pm^\alpha \pm \frac{\hbar\,J_{1,1}^\pm}{2T} \omega^\alpha\,, \\
&&	 \int_p p^\alpha p^\beta f_\pm 
=	\epsilon_\pm u^\alpha u^\beta - P_\pm \Delta^{\alpha\beta} + \pi^{\alpha\beta}_\pm
	\pm \frac{\hbar}{2}\, n_{\pm}(u^\alpha \omega^\beta + u^\beta \omega^\alpha) 
	\,,\\
&&	\int_p \frac{ p^\alpha p^\beta}{(u \cdot p)}  f_\pm
=	n_\pm u^\alpha u^\beta
	- I^\pm_{1,1}  \Delta^{\alpha\beta}
	+\mathcal{F}^{[\pi],\pm}_{1,2} \pi^{\alpha\beta}
	+ u^\alpha \nu_\pm^\beta + u^\beta \nu_\pm^\alpha
	\pm \frac{\hbar\,J_{1,1}^\pm}{2T} (u^\alpha \omega^\beta + u^\beta \omega^\alpha) \,,
\end{eqnarray}
\begin{eqnarray}
&&	\int_p \frac{ p^\alpha p^\beta p^\gamma }{(u \cdot p)}  f_\pm
=	\epsilon_\pm u^\alpha u^\beta u^\gamma
	- P_\pm \Big( u^\alpha \Delta^{\beta\gamma} + u^\beta \Delta^{\alpha\gamma}  + u^\gamma  \Delta^{\alpha\beta}\Big) 
	+ \Big( u^\alpha \pi^{\beta\gamma} + u^\beta \pi^{\alpha\gamma}  + u^\gamma  \pi^{\alpha\beta}\Big)
\nonumber\\&&\qquad\qquad\qquad\quad
	+ \frac{\mathcal{F}^{[\nu],\pm}_{1,2}}{2}
		\Big( \Delta^{\beta\gamma} \nu_\pm^{\alpha} + \Delta^{\alpha\gamma}  \nu_\pm^{\beta} + \Delta^{\alpha\beta}  \nu_\pm^{\gamma} \Big)
\nonumber\\&&\qquad\qquad\qquad\quad
	\pm \frac{\hbar\,n_\pm}{2} \Big(u^\alpha u^\beta \omega^{\gamma} + u^\alpha u^\gamma \omega^{\beta}  + u^\beta u^\gamma \omega^{\alpha} \Big)
	\mp \frac{\hbar\,J_{2,2}^\pm }{2T}
		\Big( \Delta^{\beta\gamma} \omega^{\alpha} + \Delta^{\alpha\gamma} \omega^{\beta} + \Delta^{\alpha\beta} \omega^{\gamma} \Big)
		\,,
\\
&&	\int_p \frac{ p^\alpha p^\beta p^\gamma }{(u \cdot p)^2}  f_\pm
=	n_\pm u^\alpha u^\beta u^\gamma
	- I^\pm_{1,1}  \Big( u^\alpha \Delta^{\beta\gamma} + u^\beta \Delta^{\alpha\gamma}  + u^\gamma  \Delta^{\alpha\beta}\Big) 
	+ \mathcal{F}^{[\pi],\pm}_{1,2} \Big( u^\alpha \pi^{\beta\gamma} + u^\beta \pi^{\alpha\gamma}  + u^\gamma  \pi^{\alpha\beta}\Big)
\nonumber\\&&\qquad\qquad\qquad\quad
	+\Big(u^\alpha u^\beta \nu^{\gamma}_\pm + u^\alpha u^\gamma \nu^{\beta}_\pm  + u^\beta u^\gamma \nu^{\alpha}_\pm \Big)
	+ \frac{\mathcal{F}^{[\nu],\pm}_{2,2}}{2}
		\Big( \Delta^{\beta\gamma} \nu_\pm^{\alpha} + \Delta^{\alpha\gamma}  \nu_\pm^{\beta} + \Delta^{\alpha\beta}  \nu_\pm^{\gamma} \Big)
\nonumber\\&&\qquad\qquad\qquad\quad
	\pm \frac{\hbar\,J_{1,1}^\pm}{2T}\Big(u^\alpha u^\beta \omega^{\gamma} + u^\alpha u^\gamma \omega^{\beta}  + u^\beta u^\gamma \omega^{\alpha} \Big)
	\mp \frac{\hbar\,J_{1,2}^\pm }{2T}
		\Big( \Delta^{\beta\gamma} \omega^{\alpha} + \Delta^{\alpha\gamma} \omega^{\beta} + \Delta^{\alpha\beta} \omega^{\gamma} \Big)
		\,,
\\
&&	\int_p \frac{ p^{\langle\alpha} p^{\beta\rangle} p^{\langle\gamma\rangle} p^{\langle\delta\rangle} }{(u \cdot p)^2}  f_\pm
=	2I_{2,2}^\pm \Delta^{\alpha\beta}_{\mu\nu}g^{\mu\gamma}g^{\nu\delta}
	+ \mathcal{F}^{[\pi],\pm}_{2,3}
	\Big(\frac{4}{3} g^{\rho\sigma} \Delta^{\alpha\beta}_{\mu\rho}\Delta^{\gamma\delta}_{\sigma\nu} \pi^{\mu\nu}
	+\frac{7}{9} \Delta^{\gamma\delta} \pi^{\alpha\beta}\Big)\,.
\end{eqnarray}
The following integrals of equilibrium distributions are also used:
\begin{eqnarray}
&&	\int_p \frac{ p^\alpha p^\beta p^\gamma }{(u \cdot p)^3} f_{0,\pm} 
= 	I_{0,0}^\pm  \Big[ 2u^\alpha u^\beta u^\gamma 
		- \frac{1}{3} \Big(u^\alpha g^{\beta\gamma} 
		+ u^\beta g^{\alpha\gamma} + u^\gamma g^{\alpha\beta}\Big) \Big]\,,
\\
&&	\int_p \frac{ p^\alpha p^\beta p^\gamma p^\delta}{(u \cdot p)^3} f_{0,\pm} 
= 	n_\pm \Big[ \frac{16}{5} u^\alpha u^\beta u^\gamma u^\delta 
	+ \frac{1}{15}\Big( g^{\alpha\beta}g^{\gamma\delta} + g^{\alpha\gamma}g^{\beta\delta} + g^{\alpha\delta}g^{\beta\gamma} \Big)
\nonumber\\&&\qquad\qquad
	- \frac{2}{5} \Big(u^\alpha u^\beta g^{\gamma\delta} + u^\alpha u^\gamma g^{\beta\delta} + u^\alpha u^\delta g^{\beta\gamma} 
	+ u^\beta u^\gamma g^{\alpha\delta}+ u^\beta u^\delta g^{\alpha\gamma}  + u^\gamma u^\delta g^{\alpha\beta}\Big)\Big]\,,
\\
&&	\int_p \frac{ p^\alpha p^\beta p^\gamma p^\delta}{(u \cdot p)^4} f_{0,\pm} 
= 	I_{0,0}^\pm \Big[ \frac{16}{5} u^\alpha u^\beta u^\gamma u^\delta 
	+ \frac{1}{15}\Big( g^{\alpha\beta}g^{\gamma\delta} + g^{\alpha\gamma}g^{\beta\delta} + g^{\alpha\delta}g^{\beta\gamma} \Big)
\nonumber\\&&\qquad\qquad
	- \frac{2}{5} \Big(u^\alpha u^\beta g^{\gamma\delta} + u^\alpha u^\gamma g^{\beta\delta} + u^\alpha u^\delta g^{\beta\gamma} 
	+ u^\beta u^\gamma g^{\alpha\delta}+ u^\beta u^\delta g^{\alpha\gamma}  + u^\gamma u^\delta g^{\alpha\beta}\Big)\Big] \,,
\\
&&	\int_p \frac{ p^\alpha p^\beta p^\gamma p^\delta p^\rho}{(u \cdot p)^4} f_{0,\pm} 
= 	n_\pm \Big[ \frac{16}{3} u^\alpha u^\beta u^\gamma u^\delta u^\rho
	+ \frac{1}{15}\Big(u^\alpha g^{\beta\gamma}g^{\delta\rho}  + [\text{14 other rotation terms} ]\Big)
\nonumber\\&&\qquad\qquad
	- \frac{8}{15} \Big(u^\alpha u^\beta u^\gamma g^{\delta\rho} + [\text{9 other rotation terms} ]\Big)\Big] \,.
\end{eqnarray}

\section{Collision Kernels}\label{app.collision_kernels}
In this section, we compute the collision kernels for distribution:
\begin{eqnarray}
f^{\pm} &=&  f_0^{\pm} + f_0^{\pm} (1-f_0^{\pm}) \phi_\pm[p] \,,
\\
\phi_\pm[p] &\equiv& \bigg[ 
\mp	\frac{\hbar}{2T}\frac{\omega \cdot p}{u\cdot p}
+	\lambda_{\Pi}^\pm \Pi
+	\lambda_{\nu}^\pm \nu_{\pm}^\mu p_\mu 
+	\lambda_{\pi}^\pm \pi^{\mu\nu} p_\mu p_\nu
\bigg] \,.
\end{eqnarray}
One shall keep in mind that $\lambda_\Pi$ and $\lambda_\nu$ are still functions of energy $E_p$.

Noticing that
\begin{equation}
\tilde{f}_{0,\pm}(p) = {f}_{0,\pm}(p) \cdot \exp(E_p/T - \mu^{\pm}/T)\,,
\end{equation}
one could find
\begin{eqnarray}
&&	\tilde{f}_{0,+}(p') \tilde{f}_{0,+}(k') f_{0,+}(p) f_{0,+}(k) = \tilde{f}_{0,+}(p) \tilde{f}_{0,+}(k) f_{0,+}(p') f_{0,+}(k') \,,\\
&&	\tilde{f}_{0,-}(p') \tilde{f}_{0,-}(k') f_{0,-}(p) f_{0,-}(k) = \tilde{f}_{0,-}(p) \tilde{f}_{0,-}(k) f_{0,-}(p') f_{0,-}(k') \,,\\
&&	\tilde{f}_{0,+}(p') \tilde{f}_{0,-}(k') f_{0,+}(p) f_{0,-}(k) = \tilde{f}_{0,+}(p) \tilde{f}_{0,-}(k) f_{0,+}(p') f_{0,-}(k') \,.
\end{eqnarray}

In general, we express the $\ell$-indices kernel as:
\begin{eqnarray}
	C_{+,r}^{\langle\mu_1 \cdots \mu_\ell\rangle}
&\equiv&
	\int_{\mathbf{p}} p^{\langle\mu_1} \cdots p^{\mu_\ell\rangle} E_{p}^r \,
	\mathcal{C}_{+}[f_+,f_-] \nonumber\\
&=&
	\int_{\mathbf{p}} \int_{\mathbf{p'}} \int_{\mathbf{k}} \int_{\mathbf{k'}}
	p^{\langle\mu_1} \cdots p^{\mu_\ell\rangle} E_{p}^r \nonumber\\
 &&	\quad \Big[W_{1}\big(\tilde{f}_+(p') \tilde{f}_+(k') f_+(p) f_+(k) - \tilde{f}_+(p) \tilde{f}_+(k) f_+(p') f_+(k')\big)\nonumber\\ 
 &&	\;\, +\;W_{2}\big(\tilde{f}_+(p') \tilde{f}_-(k') f_+(p) f_-(k) - \tilde{f}_+(p) \tilde{f}_-(k) f_+(p') f_-(k')\big) 
	\Big] \\
&=&
	\int_{\mathbf{p}} \int_{\mathbf{p'}} \int_{\mathbf{k}} \int_{\mathbf{k'}}
	p^{\langle\mu_1} \cdots p^{\mu_\ell\rangle} E_{p}^r \nonumber\\
 &&	\quad \Big[W_{1}\tilde{f}_{0,+}(p') \tilde{f}_{0,+}(k') f_{0,+}(p) f_{0,+}(k) \Big(\phi_+[p]+\phi_+[k] - \phi_+[p'] - \phi_+[k']\Big) \nonumber\\ 
 &&	\;\, +\;W_{2}\tilde{f}_{0,+}(p') \tilde{f}_{0,-}(k') f_{0,+}(p) f_{0,-}(k) \Big(\phi_+[p]+\phi_-[k] - \phi_+[p'] - \phi_-[k']\Big) 
	\Big] \,.
\end{eqnarray}

Then the relevant terms are
\begin{eqnarray}
	C_{+,r-1}
&=&	\Pi \int_{\mathbf{p}} \int_{\mathbf{p'}} \int_{\mathbf{k}} \int_{\mathbf{k'}} E_{p}^{r-1} \nonumber\\
 &&	\qquad \times \Big[W_{1}\tilde{f}_{0,+}(p') \tilde{f}_{0,+}(k') f_{0,+}(p) f_{0,+}(k) \Big(
 	\lambda_{\Pi}^+[E_p] - \lambda_{\Pi}^+[E_p'] + \lambda_{\Pi}^+[E_k] - \lambda_{\Pi}^+[E_k']\Big) \nonumber\\ 
 &&	\qquad \;+\; W_{2}\tilde{f}_{0,+}(p') \tilde{f}_{0,-}(k') f_{0,+}(p) f_{0,-}(k) \Big(
 	\lambda_{\Pi}^+[E_p] - \lambda_{\Pi}^+[E_p'] + \lambda_{\Pi}^-[E_k] - \lambda_{\Pi}^-[E_k']\Big) 
	\Big]
\nonumber\\
&\equiv& \mathcal{A}^{(0)}_{+,r} \, \Pi  \,, 
\end{eqnarray}
\begin{eqnarray}
	C_{+,r-1}^{\langle\mu\rangle}
&=&
	\nu_+^{\mu}\int_{\mathbf{p}} \int_{\mathbf{p'}} \int_{\mathbf{k}} \int_{\mathbf{k'}}
	\Delta^{\alpha}_{\beta} p^{\beta} E_{p}^{r-1} \nonumber\\
 &&	\qquad \times\Big[W_{1}\tilde{f}_{0,+}(p') \tilde{f}_{0,+}(k') f_{0,+}(p) f_{0,+}(k) 
 	\Big(\lambda_{\nu}^+[E_p] p_{\alpha}- \lambda_{\nu}^+[E_p']  p'_{\alpha}
	 + \lambda_{\nu}^+[E_k] k_{\alpha} - \lambda_{\nu}^+[E_k'] k'_{\alpha}\Big)\nonumber\\ 
 &&	\qquad \;+\;W_{2}\tilde{f}_{0,+}(p') \tilde{f}_{0,-}(k') f_{0,+}(p) f_{0,-}(k)
 		\Big(\lambda_{\nu}^+[E_p] p_{\alpha}- \lambda_{\nu}^+[E_p']  p'_{\alpha} \Big)
	\Big] \nonumber\\
&&+
	\nu_-^{\mu}\int_{\mathbf{p}} \int_{\mathbf{p'}} \int_{\mathbf{k}} \int_{\mathbf{k'}}
	\Delta^{\alpha}_{\beta} p^{\beta} E_{p}^{r-1} \nonumber\\
 &&	\qquad \times \Big[W_{2}\tilde{f}_{0,+}(p') \tilde{f}_{0,-}(k') f_{0,+}(p) f_{0,-}(k)
 		\Big(\lambda_{\nu}^-[E_k] k_{\alpha} - \lambda_{\nu}^-[E_k'] k'_{\alpha}\Big)
	\Big] \nonumber\\
&&+
	\frac{\hbar\,\omega^{\mu}}{2T}\int_{\mathbf{p}} \int_{\mathbf{p'}} \int_{\mathbf{k}} \int_{\mathbf{k'}}
	\Delta^{\alpha}_{\beta} p^{\beta} E_{p}^{r-1} \nonumber\\
 &&	\qquad \times \Big[W_{1}\tilde{f}_{0,+}(p') \tilde{f}_{0,+}(k') f_{0,+}(p) f_{0,+}(k) 
 	\Big(\frac{p_{\alpha}}{E_p} -\frac{p'_{\alpha}}{E_p'}
	 + \frac{k_{\alpha}}{E_k} -\frac{k'_{\alpha}}{E_k'}\Big)\nonumber\\ 
 &&	\qquad \;+\;W_{2}\tilde{f}_{0,+}(p') \tilde{f}_{0,-}(k') f_{0,+}(p) f_{0,-}(k)
 	\Big(\frac{p_{\alpha}}{E_p} - \frac{p'_{\alpha}}{E_p'}
	 - \frac{k_{\alpha}}{E_k} + \frac{k'_{\alpha}}{E_k'}\Big)
	\Big] \nonumber\\
&\equiv&
	\mathcal{A}^{(1)}_{+,r} \, \nu_+^\mu 
	+ \mathcal{B}^{(1)}_{+,r} \, \nu_-^\mu 
	+ \frac{\hbar}{2T} \mathcal{W}^{(1)}_{+,r} \,\omega^{\mu}
	\,,
\\~\nonumber\\
	C_{+,r-1}^{\langle\mu \nu\rangle}
&=&
	\pi^{\mu\nu}\int_{\mathbf{p}} \int_{\mathbf{p'}} \int_{\mathbf{k}} \int_{\mathbf{k'}}
	\Delta_{\alpha'\beta'}^{\alpha\beta}p^{\alpha'}  p^{\beta'} E_{p}^{r-1} \nonumber\\
 &&	\qquad \times \Big[W_{1}\tilde{f}_{0,+}(p') \tilde{f}_{0,+}(k') f_{0,+}(p) f_{0,+}(k) 
 	\frac{p_{\alpha} p_{\beta}- p'_{\alpha} p'_{\beta} 
	 + k_{\alpha} k_{\beta} - k'_{\alpha} k'_{\beta}}{4J_{4,2}^+}\nonumber\\ 
 &&	\qquad \;+\;W_{2}\tilde{f}_{0,+}(p') \tilde{f}_{0,-}(k') f_{0,+}(p) f_{0,-}(k)
 	\Big(\frac{p_{\alpha} p_{\beta} - p'_{\alpha} p'_{\beta}}{4J_{4,2}^+} 
 	+\frac{k_{\alpha} k_{\beta}  - k'_{\alpha} k'_{\beta}}{4J_{4,2}^-} \Big)
	\Big] 
\nonumber\\
&\equiv& \mathcal{A}^{(2)}_{+,r} \, \pi^{\mu\nu}  \,. 
\end{eqnarray}

The following term is also needed:
\begin{eqnarray}
&&	\frac{\hbar}{2} \omega^\gamma \Delta^{\mu\nu}_{\alpha\beta} \int_p p^\alpha p^\beta p_\gamma E_p^{r-1} \mathcal{C}_+
\nonumber\\&=& 
	\frac{\hbar}{2}  \Delta^{\mu\nu}_{\rho\sigma} \omega^\rho \nu_+^\sigma  \cdot \frac{2}{15}\int_p \int_{p'} \int_{k} \int_{k'} 
	(\Delta^{\alpha'\beta'} p_{\alpha'} p_{\beta'}) \Delta^{\alpha\beta} p_\beta E_p^{r-1}
\nonumber\\  &&
	\qquad \times \Big[ W_{1}\tilde{f}_{0,+}(p') \tilde{f}_{0,+}(k') f_{0,+}(p) f_{0,+}(k) 
 	\Big(\lambda_{\nu}^+[E_p] p_{\alpha}- \lambda_{\nu}^+[E_p']  p'_{\alpha}
	 + \lambda_{\nu}^+[E_k] k_{\alpha} - \lambda_{\nu}^+[E_k'] k'_{\alpha}\Big)
\nonumber\\  &&
	\qquad \;+\; W_{2}\tilde{f}_{0,+}(p') \tilde{f}_{0,-}(k') f_{0,+}(p) f_{0,-}(k)
 		\Big(\lambda_{\nu}^+[E_p] p_{\alpha}- \lambda_{\nu}^+[E_p']  p'_{\alpha} \Big)\Big]
\nonumber\\  &&
	+\frac{\hbar}{2}  \Delta^{\mu\nu}_{\rho\sigma} \omega^\rho \nu_-^\sigma  \cdot \frac{2}{15}\int_p \int_{p'} \int_{k} \int_{k'} 
	(\Delta^{\alpha'\beta'} p_{\alpha'} p_{\beta'}) \Delta^{\alpha\beta} p_\beta E_p^{r-1}
\nonumber\\  &&
	\qquad \times \Big[ W_{2}\tilde{f}_{0,+}(p') \tilde{f}_{0,-}(k') f_{0,+}(p) f_{0,-}(k)
 		\Big(\lambda_{\nu}^-[E_k] k_{\alpha} - \lambda_{\nu}^-[E_k'] k'_{\alpha}\Big)\Big]
\nonumber\\  &\equiv&
	  \mathcal{X}_{2,r}^{+,+} \frac{\hbar}{2}  \Delta^{\mu\nu}_{\rho\sigma} \omega^\rho \nu_+^\sigma
	+ \mathcal{X}_{2,r}^{+,-} \frac{\hbar}{2}  \Delta^{\mu\nu}_{\rho\sigma} \omega^\rho \nu_-^\sigma \,.
\end{eqnarray}

\end{widetext}
\end{appendix}

\end{document}